 \documentclass[aps,twocolumn,pra,superscriptaddress,amsmath,showpacs,tightenlines]{revtex4}

\usepackage{epsfig,graphicx,times}
\usepackage{amstext}
\usepackage{amsmath}            %serve per le subequazioni
\usepackage{amssymb}            %serve per il simbolo "marchio registrato", \circledR
\usepackage{graphicx}           %serve per le figure eps, ps etcyy
\usepackage{latexsym}
\usepackage{bm}

\usepackage{color}

\newtheorem{criterion}{Criterion}
\newtheorem{observation}{Observation}

\newcommand{\tr}{{\rm Tr}\,}
\def\<{\langle}
\def\>{\rangle}

\def\A{\hat a_1^\dagger}
\def\a{\hat a_1}
\def\B{\hat a_2^\dagger}
\def\b{\hat a_2}
\def\C{\hat a_3^\dagger}
\def\c{\hat a_3}
\def\f{\hat f}
\def\F{\hat F}
\def\na{\<\hat n_1\>}
\def\nb{\<\hat n_2\>}
\def\mn{\<\hat n_1 \hat n_2\>}

\def\varn#1{{\langle :(\Delta \hat #1)^2: \rangle}}
\def\var#1{{\langle (\Delta \hat #1)^2 \rangle}}

\def\normal#1{{\langle :  #1 : \rangle}}
\def\normalo#1{{\langle \dd  #1 \dd \rangle}}
\def \dfn{{d_{\hat F}^{{\rm (n)}}}}
\newcommand\dfnnew[1]{{d_{#1}^{{\rm (n)}}}}
\def \dn{{d^{\rm (n)}}}
\def \dPT{{d^{\Gamma}}}
\def \intda{{\int {\rm d}^2\bm{\alpha}\;}}

\def \dfPT{{d_{\hat F}^{\Gamma}}}
\def\mean#1{{\langle #1 \rangle}}

\newcommand{\dd}{{\rm\raisebox{-2.pt}{$\stackrel{\scriptstyle\circ}{\scriptstyle\circ}$}}}
\newcommand{\ncl}{{\;\rm\raisebox{-2.pt}{\,$\stackrel{\rm ncl}{<}\;$}}}
\newcommand{\cl}{{\;\rm\raisebox{-2.pt}{$\stackrel{\rm cl}{\ge}\;$}}}
\newcommand{\ent}{{\;\rm\raisebox{-2.pt}{\,$\stackrel{\rm ent}{<}\;$}}}

\newcommand\DET[2]{{\left|\begin{array}{cc} #1 \\ #2 \end{array} \right|}}
\newcommand\DETT[3]{{\left|\begin{array}{ccc} #1 \\ #2  \\ #3 \end{array} \right|}}
\newcommand\Det[4]{{\left|\begin{array}{cccc} #1 \\ #2  \\ #3  \\ #4 \end{array} \right|}}
\newcommand\Mat[4]{{\left(\begin{array}{cccc} #1 \\ #2  \\ #3  \\ #4 \end{array} \right)}}

\def\extra#1{{\em #1}}

\begin{document}

\title{Testing nonclassicality in multimode fields:\\
a unified derivation of classical inequalities}

\author{Adam Miranowicz}
\affiliation{Advanced Science Institute, RIKEN, Wako-shi, Saitama
351-0198, Japan} \affiliation{Faculty of Physics, Adam Mickiewicz
University, PL-61-614 Pozna\'n, Poland}

\author{Monika Bartkowiak}
\affiliation{Faculty of Physics, Adam Mickiewicz University,
PL-61-614 Pozna\'n, Poland}

\author{Xiaoguang Wang}
\affiliation{Advanced Science Institute, RIKEN, Wako-shi, Saitama
351-0198, Japan} \affiliation{Zhejiang Institute of Modern
Physics, Department of Physics, Zhejiang University, Hangzhou
310027, China}

\author{Yu-xi Liu}
\affiliation{Institute of Microelectronics, Tsinghua University,
Beijing 100084, China} \affiliation{Tsinghua National Laboratory
for Information Science and Technology (TNList), Tsinghua
University, Beijing 100084, China} \affiliation{Advanced Science
Institute, RIKEN, Wako-shi, Saitama 351-0198, Japan}

\author{Franco Nori}
\affiliation{Advanced Science Institute, RIKEN, Wako-shi, Saitama
351-0198, Japan} \affiliation{Physics Department, The University
of Michigan, Ann Arbor, Michigan 48109-1040, USA}

\date{\today}

\begin{abstract}
We consider a way to generate operational inequalities to test
nonclassicality (or quantumness) of multimode bosonic fields (or
multiparty bosonic systems) that unifies the derivation of many
known inequalities and allows to propose new ones. The
nonclassicality criteria are based on Vogel's criterion
corresponding to analyzing the positivity of multimode
$P$~functions or, equivalently, the positivity of matrices of
expectation values of, e.g., creation and annihilation operators.
We analyze not only monomials, but also polynomial functions of
such moments, which can sometimes enable simpler derivations of
physically relevant inequalities. As an example, we derive various
classical inequalities which can be violated only by nonclassical
fields. In particular, we show how the criteria introduced here
easily reduce to the well-known inequalities describing: (a)
multimode quadrature squeezing and its generalizations including
sum, difference and principal squeezing, (b) two-mode one-time
photon-number correlations including sub-Poisson photon-number
correlations and effects corresponding to violations of the
Cauchy-Schwarz and Muirhead inequalities, (c) two-time single-mode
photon-number correlations including photon antibunching and
hyperbunching, and (d) two- and three-mode quantum entanglement.
Other simple inequalities for testing nonclassicality are also
proposed. We have found some general relations between the
nonclassicality and entanglement criteria, in particular, those
resulting from the Cauchy-Schwarz inequality. It is shown that
some known entanglement inequalities can be derived as
nonclassicality inequalities within our formalism, while some
other known entanglement inequalities can be seen as sums of more
than one inequality derived from the nonclassicality criterion.
This approach enables a deeper analysis of the entanglement for a
given nonclassicality.
\end{abstract}


% 42.50.Ar Photon statistics and coherence theory
% 42.50.Xa for optical tests of quantum theory
% 03.67.Mn Entanglement measures, witnesses, and other characterizations

% 42.50.Dv Quantum state engineering and measurements in quantum optics
% 03.65.Wj State reconstruction, quantum tomography
% 03.65.Ta Foundations of quantum mechanics; measurement theory
% 03.65.Ud Entanglement and quantum nonlocality
%         (e.g. EPR paradox, Bell's inequalities, GHZ states, etc.)

\maketitle \pagenumbering{arabic}

%------------------------------------------------------------------
\section{Introduction}

Testing whether a given state of a system cannot be described
within a classical theory, has been one of the fundamental
problems of quantum theory from its beginnings to current studies
in, e.g., quantum
optics~\cite{Glauber,Sudarshan,DodonovBook,VogelBook,MandelBook,PerinaBook,Walls79,Loudon80,Loudon87,Smirnov87,Klyshko96,Dodonov02},
condensed matter (see, e.g., Refs.~\cite{DodonovBook,Nori}),
nanomechanics~\cite{Schwab05,Wei06}, and quantum biology (see,
e.g., Ref.~\cite{qbiology}). Macroscopic quantum superpositions
(being at the heart of the Schr\"odinger-cat paradox) and related
entangled states (which are at the core of the
Einstein-Podolsky-Rosen paradox and Bell's theorem) are famous
examples of nonclassical states which are not only physical
curiosities but now fundamental resources for quantum-information
processing~\cite{Nielsen}.

All states (or phenomena) are quantum, i.e., nonclassical. Thus,
it is quite arbitrary to call some states ``classical''.
Nevertheless, some states are closer to their classical
approximation than other states. The most classical {\em pure}
states of the harmonic oscillator are coherent states. Thus,
usually, they are considered classical, while all other pure
states of the harmonic oscillator are deemed nonclassical. The
nonclassicality criterion for {\em mixed} states is more
complicated and it is based on the Glauber-Sudarshan
$P$~function~\cite{Glauber,Sudarshan}. A commonly accepted formal
criterion which enables to distinguish nonclassical from classical
states reads as
follows~\cite{DodonovBook,VogelBook,MandelBook,PerinaBook}: A
quantum state is {\em nonclassical} if its Glauber-Sudarshan $P$
function cannot be interpreted as a {\em true} probability
density. Note that, according to this definition, any entangled
state is nonclassical, but not every separable state is classical.

Various operational criteria of nonclassicality (or quantumness)
of single-mode fields were proposed~(see, e.g.,
Refs.~\cite{DodonovBook,VogelBook,Richter02,Rivas} and references
therein). In particular, Agarwal and Tara~\cite{Agarwal92},
Shchukin, Richter and Vogel (SRV)~\cite{NCL1,NCL2} proposed
nonclassicality criteria based on matrices of moments of
annihilation and creation operators for single-mode fields.
Moreover, an efficient method for measuring such moments was also
developed by Shchukin and Vogel~\cite{SVdetect}.

It is not always sufficient to analyze a single-mode field, i.e.,
an elementary excitation of a normal mode of the field confined to
a one-dimensional cavity. To describe the generation or
interaction of two or more bosonic fields, the standard analysis
of single-system nonclassicality should be generalized to the two-
and multi-system (multimode)  case. Simple examples of such
bosonic fields are multimode number states, multimode coherent and
squeezed light, or fields generated in multi-wave mixing,
multimode scattering, or multi-photon resonance.

Here, we study in greater detail and modify an operational
criterion of nonclassicality for multimode radiation fields of
Vogel \cite{Vogel08}, which is a generalized version of the SRV
nonclassicality criterion~\cite{NCL1,NCL2} for single-mode fields.
It not only describes the multimode fields, but can also be
applied in the analysis of the dynamics of radiation sources. This
could be important for the study of, e.g., time-dependent
correlation functions, which are related to time-dependent field
commutation rules (see, e.g., subsections 2.7 and 2.8 in
Ref.~\cite{VogelBook}).

A variety of multimode nonclassicality inequalities has been
proposed in quantum optics~(see, e.g.,
textbooks~\cite{DodonovBook,VogelBook,MandelBook,PerinaBook},
reviews~\cite{Walls79,Loudon80,Loudon87,Klyshko96,Smirnov87}, and
Refs.~\cite{Yuen76,Kozierowski77,Caves85,Reid86,Dalton86,Schleich87,Agarwal88,Luks88,Hillery89,Lee90,Zou90,Klyshko96pla,Miranowicz99a,Miranowicz99b,An99,An00,Jakob01})
and tested experimentally (see, e.g.,
Refs.~\cite{Clauser74,Kimble77,Short83,Slusher85,Grangier86,Hong87,Lvovsky02}).
The nonclassicality criterion described here enables a simple
derivation of them. Moreover, it offers an effective way to derive
new inequalities, which might be useful in testing the
nonclassicality of specific states generated in experiments.

It is worth noting that we are analyzing nonclassicality criteria
but {\em not} a degree of nonclassicality. We admit that the
latter problem is experimentally important and a few ``measures''
of nonclassicality have been
proposed~\cite{Hillery87,Lee92,Lutkenhaus95,Dodonov00,Marian02,Dodonov03,Malbouisson03,Kenfack04,Asboth05,Boca09}.

Analogously to the SRV nonclassicality criteria, Shchukin and
Vogel~\cite{SV05} proposed an entanglement criterion based on the
matrices of moments and partial transposition. This criterion was
later amended~\cite{MP06} and generalized~\cite{MPHH09} to replace
partial transposition by nondecomposable positive maps and
contraction maps (e.g., realignment). A similar approach for
entanglement verification, based on the construction of matrices
of expectation values, was also investigated in
Refs.~\cite{Rigas06,Korbicz06,Moroder08,Haseler08}.

Here we analyze relations between classical inequalities derived
from the two- and three-mode nonclassicality criteria and the
above-mentioned entanglement criterion.

The article is organized as follows: In Sect.~\ref{Sect2}, a
nonclassicality criterion for multimode bosonic fields is
formulated. We apply the criterion to rederive known and a few
apparently new nonclassicality inequalities. In
subsection~\ref{Sect3a}, we summarize the Shchukin-Vogel
entanglement criterion~\cite{SV05,MP06}. In
subsection~\ref{Sect3b}, we apply it to show that some known
entanglement inequalities (including those of Duan {\em et
al.}~\cite{Duan} and Hillery and Zubairy~\cite{Hillery06}) exactly
correspond to unique nonclassicality inequalities. In
subsection~\ref{Sect3c}, we analyze such entanglement inequalities
(including Simon's criterion~\cite{Simon}) that are represented
apparently not by a single inequality but by sums of inequalities
derived from the nonclassicality criterion. Moreover, other
entanglement inequalities are derived in subsection~\ref{Sect3c2}.
The discussed nonclassicality and entanglement criteria are
summarized in Tables~I and~II. We conclude in Sect.~\ref{Sect4}.

%------------------------------------------------------------------
\section{Nonclassicality criteria for multimode fields\label{Sect2}}

An $M$-mode bosonic state $\hat\rho$ can be completely described
by the Glauber-Sudarshan $P$~function defined
by~\cite{Glauber,Sudarshan}:
\begin{eqnarray}
  \hat\rho &=& \intda P(\bm{\alpha,\alpha}^*)|\bm{\alpha}\>\< \bm{\alpha}|,
  \label{N01}
\end{eqnarray}
where $|\bm{\alpha}\>= \prod_{m=1}^M|\alpha_m\>$ and $|\alpha_m\>$
is the $m$th-mode coherent state, i.e., the eigenstate of the
$m$th-mode annihilation operator $\hat a_m$, $\bm{\alpha}$ denotes
complex multivariable $(\alpha_1,\alpha_2,...,\alpha_M)$, and
${\rm d}^2 \bm{\alpha}=\prod_{m}{\rm d}^2\alpha_m$. The density
matrix $\hat\rho$ can be supported on the tensor product of either
infinite-dimensional or finite-dimensional Hilbert spaces. For the
sake of simplicity, we assume the number $M$ of modes to be
finite. But there is no problem to generalize our results for an
infinite number of modes.

A criterion of nonclassicality is usually formulated as
follows~\cite{Titulaer65}:
\begin{criterion} %1
A multimode bosonic state $\hat\rho$ is considered to be
nonclassical if its Glauber-Sudarshan $P$~function cannot be
interpreted as a classical probability density, i.e., it is
nonpositive or more singular than Dirac's delta function.
Conversely, a state is called classical if it is described by a
$P$~function being a classical probability density.
\end{criterion}

It is worth noting that Criterion~1 (and the following criteria)
does not have a fundamental indisputable validity, and it was the
subject of criticism by, e.g., W\"unsche~\cite{Wunsche04}, who
made the following two observations. (i) In the vicinity of any
classical state there are nonclassical states, as can be
illustrated by analyzing modified thermal states. So, arbitrarily
close to any classical state there is a nonclassical state giving,
to arbitrary precision, exactly the same outcomes as for the
classical state in any measurement. Note that analogous problems
can be raised for entanglement criteria~\cite{MPHH09} for
continuous-variable systems, as in the vicinity of any separable
state there are entangled states.~\footnote{It is worth stressing
that this is the case only for continuous-variable systems: in the
finite dimensional case, the set of separable states has finite
volume.}  (ii) There are intermediate quasiclassical (or
unorthodox classical) states, which {\em cannot} be clearly
classified as classical or nonclassical according to Criterion~1.
This can be illustrated by analyzing the squeezing of thermal
states, which does not lead immediately from classical to
nonclassical states.

Due to the singularity of the $P$~function, Criterion~1 is not
operationally useful as it is extremely difficult (although
sometimes possible~\cite{Kiesel}) to directly reconstruct the
$P$~function from experimental data.

Recently, Shchukin, Richter and Vogel~\cite{NCL1,NCL2} proposed a
hierarchy of operational criteria of nonclassicality of
single-mode bosonic states. This approach is based on the normally
ordered moments of, e.g., annihilation and creation operators or
position and momentum operators. An infinite set of these criteria
(by inclusion of the correction analogous to that given in
Ref.~\cite{MP06}) corresponds to a single-mode version of
Criterion~1.

Let us consider a (possibly infinite) countable set
$\F=(\f_{1},\f_{2},\ldots,\f_{i},\ldots)$ of  $M$-mode operators
$\f_i\equiv \f_i (\hat {\bf a},\hat {\bf a}^\dagger)$, each a
function of annihilation, $\hat {\bf a}\equiv (\hat a_1,\hat
a_2,...,\hat a_M)$, and creation, $\hat {\bf a}^\dagger$,
operators. For example, we may choose such operators as monomials
\begin{eqnarray}
\label{eq:product}
  \f_{i}
  = \prod_{m=1}^M (\hat a^\dagger_m)^{i_{2m-1}} \hat a_m^{i_{2m}},
\end{eqnarray}
where $i$ stands in this case for the multi-index ${\bf i}\equiv
(i_{1},i_{2},...,i_{2M})$, but the  $\f_{i}$'s can be more
complicated functions, for example polynomials in the creation and
annihilation operators.

If
\begin{equation}
\label{N02}
  \f
  = \sum_{i} c_{i} \f_{i},
\end{equation}
where $c_{i}$ are arbitrary complex numbers, then with the help of
the $P$~function one can directly calculate the normally ordered
(denoted by $::$) mean values of the Hermitian operator
$\f^\dagger \f$ as follows~\cite{NCL1,Korbicz}:
\begin{eqnarray}
  \normal{\f^\dagger \f } &=& \intda |f(\bm{\alpha,\alpha}^*)|^2
  P(\bm{\alpha,\alpha}^*) .
  \label{N03}
\end{eqnarray}
The crucial observation of SRV~\cite{NCL1} in the derivation of
their criterion is the following:

\begin{observation}
\label{obs:obsSRV} If the $P$~function for a given state is a
classical probability density, then ${\normal{\f^\dagger \f}}\ge
0$ for any function $\f$. Conversely, if $\normal{ \f^\dagger \f }
< 0$ for some $\f$, then the $P$~function is not a classical
probability density.
\end{observation}
The condition based on nonpositivity of the $P$~function is
usually considered a necessary and sufficient condition of
nonclassicality. In fact, as shown by Sperling~\cite{Sperling}, if
the $P$~function is more singular than Dirac's $\delta$-function
[e.g., given by the $n$th derivative of $\delta(\alpha)$ for
$n=1,2,...$], then it is also nonpositive.

With the help of Eq.~(\ref{N02}), $\normal{\f^\dagger \f}$ can be
given by
\begin{eqnarray}
  \normal{\f^\dagger \f} &=& \sum_{i,j} c^*_{i}c_{j}
  M^{\rm (n)}_{ij}(\hat\rho)
\label{N04}
\end{eqnarray}
in terms of the normally ordered correlation functions
\begin{eqnarray}
  M^{\rm (n)}_{ij}(\hat\rho) &=& \tr(: \f_{i}^\dagger
  \f_{j}:\, \hat\rho),
\label{N05}
\end{eqnarray}
where the superscript $(n)$ (similarly to $:\,:$) denotes the
normal order of field operators. In the special case of two modes,
analyzed in detail in the next sections, and with the choice
of $\f_i$ given by Eq.~\eqref{eq:product}, Eq.~(\ref{N05})
can be simply written as
\begin{equation}
  M^{\rm (n)}_{ij}(\hat\rho)
% \equiv M_{i_1i_2i_3i_4,j_1j_2j_3j_4}(\hat\rho)
=\tr\big[:({\hat a}^{\dagger i_{1}}{\hat a}^{i_{2}}{\hat
b}^{\dagger i_{3}}{\hat b}^{i_{4}})^\dagger ({\hat a}^{\dagger
j_{1}}{\hat a}^{j_{2}}{\hat b} ^{\dagger j_{3}}{\hat
b}^{j_{4}}):\hat\rho\big], \label{N06}
\end{equation}
where $\hat a=\hat a_1$ and $\hat b=\hat a_2$. It is worth noting
that there is an efficient optical scheme \cite{SVdetect} for
measuring the correlation functions~(\ref{N06}).

With a set $\F=(\f_{1},\f_{2},\ldots,\f_{i},\ldots)$ fixed, the
correlations \eqref{N05} form a (possibly infinite) Hermitian
matrix
\begin{eqnarray}
  M^{\rm (n)}(\hat\rho)
  = [M^{\rm (n)}_{ij}(\hat\rho)].
  \label{N07}
\end{eqnarray}
In order to emphasize the dependence of~(\ref{N07}) on the choice
of $\F$, we may write  $M^{\rm (n)}_{\F}(\hat\rho)$. Moreover, let
$[M^{\rm (n)}(\hat\rho)]_{\bf r}$, with ${\bf
r}=(r_1,\ldots,r_N)$, denote the $N \times N$ principal submatrix
of $M^{\rm (n)}(\hat\rho)$ obtained by deleting all rows and
columns except the ones labeled by $r_1,\ldots,r_N$.

Analogously to Vogel's approach \cite{Vogel08}, by applying
Sylvester's criterion (see, e.g., \cite{Strang,MP06}) to the
matrix~(\ref{N07}), a generalization of the single-mode SRV
criterion for multimode fields can be formulated as follows:
\begin{criterion} %2
For any choice of $\F=(\f_{1},\f_{2},\ldots,\f_{i},\ldots)$, a
multimode state $\hat\rho$ is nonclassical if there exists a
negative principal minor, i.e., $\det [M_{\F}^{\rm
(n)}(\hat\rho)]_{\bf r}< 0$, for some ${\bf
r}\equiv(r_1,\ldots,r_N)$, with $1\le r_1< r_2<\ldots < r_{N}$.
\end{criterion}
According to Vogel~\cite{Vogel08}, this criterion (and the
following Criterion~3) can also be applied to describe the
nonclassicality of space-time correlations and the dynamics of
radiation sources by applying the generalized $P$~function:
\begin{eqnarray}
  P(\bm{\alpha,\alpha}^*) &=& \left\langle \dd \prod_{i=1}^M
  \delta(\hat a_i - \alpha_i)\dd \right\rangle.
\label{VogelP}
\end{eqnarray}
where $\bm{\alpha}=(\alpha_1,...,\alpha_M)$, with
$\alpha_i=\alpha_i({\bf r}_i,t_i)$ depending on the space-time
arguments $({\bf r}_i,t_i)$.  By contrast to the standard
definition of $P$~function, symbol $\dd\,\dd$ describes both the
normal order of field operators and also time order, i.e., time
arguments increase to the right (left) in products of creation
(annihilation) operators~\cite{VogelBook}. As an example, we will
apply this generalized criterion to show the nonclassicality of
photon antibunching and hyperbunching effects in Appendix C.

Note that Criterion~2, even for the choice of $\f_i$ given by
Eq.~\eqref{eq:product} and in the special case of single-mode
fields, does not exactly reduce to the SRV criterion as it
appeared in Ref.~\cite{NCL2}. To show this, let us denote by
$M^{\rm (n)}_N(\hat\rho)$  the submatrix corresponding to the
first $N$ rows and columns of $M^{\rm (n)}(\hat\rho)$. According
to the original SRV criterion (Theorem~3 in Ref.~\cite{NCL2}), a
single-mode state is nonclassical if there exists an $N$, such
that the leading principal minor is negative, i.e. $\det M^{\rm
(n)}_N(\hat\rho)<0$. Such formulated criterion fails for singular
(i.e., $\det M^{\rm (n)}_N(\hat\rho) =0$) matrices of moments, as
explained in detail in the context of quantum entanglement in
Ref.~\cite{MP06}.

Considering $[M^{\rm (n)}_{\F}(\hat\rho)]_{\bf r}$ is equivalent
to considering the correlation matrix corresponding to a subset
$\F'\subset \F$, with $\F'=(\f_{r_1},\f_{r_2},...,\f_{r_N})$,
i.e., $[M^{\rm (n)}_{\F}(\hat\rho)]_{\bf r}=M^{\rm
(n)}_{\F'}(\hat\rho)$. We note that the subset symbol is used for
brevity although it is not very precise, as the $\F$s are ordered
collections of operators.

Thus, by denoting
\begin{eqnarray}
 M^{\rm (n)}_{\F'}(\hat\rho)  \equiv
 [M^{\rm (n)}_{\F}(\hat\rho)]_{\bf r} \hspace{4.5cm}
\nonumber\\
= \Mat{ \normal{\f_{r_1}^\dagger \f_{r_1}} &
\normal{\f_{r_1}^\dagger \f_{r_2}} & \cdots &
\normal{\f_{r_1}^\dagger \f_{r_N}} } { \normal{\f_{r_2}^\dagger
\f_{r_1}} & \normal{\f_{r_2}^\dagger \f_{r_2}} & \cdots &
\normal{\f_{r_2}^\dagger \f_{r_N}} } { \vdots & \vdots & \ddots &
\vdots } { \normal{\f_{r_N}^\dagger \f_{r_1}} &
\normal{\f_{r_N}^\dagger \f_{r_2}} & \cdots &
\normal{\f_{r_N}^\dagger \f_{r_N}} },
  \label{N08}
\end{eqnarray}
and its determinant
\begin{eqnarray}
\dfnnew{\F'}(\hat\rho)\equiv \det \,
  M^{\rm (n)}_{\F'}(\hat\rho),
  %\equiv \det \, (M^{\rm (n)}_{\F}(\hat\rho))_{\bf r}
  \label{N08a}
\end{eqnarray}
we can equivalently rewrite Criterion~2 as:
\begin{criterion} %3
A multimode bosonic state $\hat\rho$ is nonclassical if there
exists $\F$, such that $\dfnnew{\F}(\hat\rho)$ is negative.
\end{criterion}
This can be written more compactly as:
\begin{eqnarray}
 \hat\rho \textrm{~is classical} &\Rightarrow& \forall {
\F}: \quad \dfn(\hat\rho) \ge 0,
\nonumber \\
 \hat\rho \textrm{~is nonclassical} &\Leftarrow& \exists
{\F}: \quad \dfn(\hat\rho) <0. \label{N09}
\end{eqnarray}
In the following, we use the symbol $\ncl$ to emphasize that a
given inequality {\em can} be satisfied only for {\em
nonclassical} states and the symbol $\cl$ to indicate that an
inequality {\em must} be satisfied for all {\em classical} states.

Let us comment further on the relation between Criteria 2 and 3
and the SRV criterion (in its amended version that takes into
account the issue of singular matrices). Criterion 3 corresponds
to checking the positivity of an infinite matrix $M_{ij}^{(n)}$
defined as in \eqref{N05} with the $\f_i$'s chosen to be all
possible monomials given by Eq.~\eqref{eq:product}. Considering
the positivity of larger and larger submatrices of this matrix
leads to a hierarchy of criteria: testing the positivity of some
submatrix $M^{(n)}_N$ leads to a stronger criterion than testing
the positivity of a submatrix $M^{(n)}_{N'}$, with $N'< N$.
Nonetheless, when one invokes Sylvester's criterion in order to
transform the test of positivity of a matrix into the test of
positivity of its many principal minors, it is arguably difficult
to speak of a ``hierarchy''. Indeed, because of the issue of the
possible singularity of the matrix we cannot simply consider,
e.g., leading principal minors involving larger and larger
submatrices.

As regards the general formalism, of course by adding operators to
the set $\F$, and therefore increasing the dimension of the matrix
$M^{(n)}_{\F'}$, one obtains a hierarchy of \emph{matrix}
conditions on classicality. Nonetheless, also in our case when
moving to scalar inequalities by considering determinants, we face
the issue of the possible singularity of matrices. Motivated also
by this difficulty, in the present article we do not focus so much
on the idea a hierarchy of criteria, but rather explore the
approach that by using matrices of expectations values it is
possible to easily obtain criteria of nonclassicality and
entanglement in the form of inequalities. As already explained,
this is done by referring to Observation \ref{obs:obsSRV} and
considering $\f_i$'s possibly more general than monomials, e.g.,
polynomials.

Indeed, when we choose a set of operators $\F=(\f_1,\f_2,\dots)$,
we compute the corresponding matrix of expectation values, and we
check its positivity, what we are doing is equivalent to checking
positivity of, e.g., ${\normal{\f^\dagger \f}}$ for all $f$'s that
can be written as a linear combination of the operators in $\F$:
$\f=\sum_ic_i\f_i$. As polynomials can be expanded into monomials,
it is clear that checking the positivity of a matrix
$M^{(n)}_{\F}$ with $\F$ consisting of polynomials, cannot give a
stronger criterion than checking the positivity of a matrix
$M^{(n)}_{\F'}$, where $\F'$ is given by all the monomials
appearing in the elements of $\F$. Of course, to have a stronger
\emph{matrix} criterion of classicality we pay a price in terms of
the dimension of the matrix $M^{(n)}_{\F'}$, which is larger than
$M^{(n)}_{\F}$. Further, as we will see, by considering general
sets $\F$---that is, not only containing monomials---one can
straightforwardly obtain interesting and physically relevant
inequalities, which may be difficult to pinpoint when considering
monomials as ``building blocks''. It is worth noting that the
possibility of using polynomial functions of moments was also
discussed in Ref.~\cite{SV05} in the context of entanglement
criterion.

Finally, we remark that to make the above criteria sensitive in
detecting nonclassicality, the $f_i$ must be chosen such that the
normal ordering is important in giving $M^{(n)}$. In particular,
assuming this special structure for the $f_i$'s, there must be
some combination of creation and annihilation operators. On the
contrary, the inclusion of only creation or only annihilation
operators would give a matrix $M^{(n)}$ positive for every state,
thus completely useless for detecting nonclassicality.

%------------------------------------------------------------------
\begin{table*}[tbp]
\caption{Criteria for single-time nonclassical effects in two-mode
(TM) and multimode (MM) fields, and two-time nonclassical effects
in single-mode (SM) fields.}
\begin{center}
\begin{tabular}{l c c}
\hline\hline {\bf Nonclassical effect} & {\bf Criterion} &
Equations
\\ \hline\hline
MM quadrature squeezing & $\dn(1,\hat X_{\bm{\phi}})<0$ & (\ref{N10}),~(\ref{N15})\\[5pt]
TM principal squeezing of Luk\v{s} {\em et al.}~\cite{Luks88} & $\dn(\Delta \hat a_{12}^\dagger,\Delta \hat a_{12})=\dn(1, \hat a_{12}^\dagger, \hat a_{12})<0$ & (\ref{N16})--(\ref{z36}) \\[5pt]
TM sum squeezing of Hillery~\cite{Hillery89} & $\dn(1,\hat V_{\phi})<0$ & (\ref{N20}),~(\ref{N22}) \\[5pt]
MM sum  squeezing of An-Tinh~\cite{An99} & $\dn(1,\hat {\cal V}_{\phi})<0$ & (\ref{z9}),~(\ref{z10})\\[5pt]
TM difference squeezing of Hillery~\cite{Hillery89} & $\dn(1,\hat W_{\phi})<- \frac12 \min \left(\mean{\hat n_1},\mean{\hat n_2}\right)$ & (\ref{N23}),~(\ref{N26}),~(\ref{z1})\\[5pt]
MM difference squeezing of An-Tinh~\cite{An00}& $\dn(1,\hat {\cal W}_{\phi})<-\frac14 \left||\mean{\hat C}|-\mean{\hat D}\right|$ & (\ref{z18}),~(\ref{z19}) \\[5pt]
TM sub-Poisson photon-number correlations & $\dn(1,\hat n_1 \pm\hat n_2)<0$ & (\ref{N28}),~(\ref{N30}) \\[5pt]
Cauchy-Schwarz inequality violation& $\dn(\f_1,\f_2)<0$ & (\ref{x94}),~(\ref{x96}) \\[5pt]
TM Cauchy-Schwarz inequality violation via Agarwal's test~\cite{Agarwal88}& $\dn(\hat n_1,\hat n_2)<0$ & (\ref{x15}),~(\ref{x17}) \\[5pt]
TM Muirhead inequality violation via Lee's test~\cite{Lee90}& $\dn(\hat n_1-\hat n_2)<0$ & (\ref{x30}),~(\ref{x30a}) \\[5pt]
SM photon antibunching& $\dn[\hat n(t),\hat n(t+\tau)]<0$ & (\ref{y05}),~(\ref{x23}) \\[5pt]
SM photon hyperbunching& $\dn[\Delta\hat n(t),\Delta\hat n(t+\tau)]$ & (\ref{y05a}),~(\ref{x27}),~(\ref{z34}) \\
&\quad $=\dn[1,\hat n(t),\hat n(t+\tau)]<0$ &  \\[4pt]
Other TM nonclassical effects
& $\dn(1,\a\b,\A\B)<0$ & (\ref{x72}) \\[5pt]
& $\dn(1,\a\B,\A \b)<0$ & (\ref{x78}) \\[5pt]
& $\dn(1,\a+\B,\A +\b)<0$ & (\ref{x84}) \\[5pt]
& $\dn(1,\a+\b,\A +\B)<0$ & (\ref{x90}) \\[5pt]
& $\dn(1,\a,\A,\B,\b)<0$ & (\ref{x36}) \\[2pt] \hline \hline
\end{tabular}
\end{center}
\end{table*}
%------------------------------------------------------------------
\begin{table*}[tbp]
\caption{Entanglement criteria via nonclassicality criteria.}
\begin{center}
\begin{tabular}{l c c c}
\hline\hline Reference & {\bf Entanglement criterion} & {\bf
Equivalent nonclassicality criterion} & Equations
\\ \hline\hline
Duan {\em et al.}~\cite{Duan}        & $\dPT(\Delta\a,\Delta\b)=\dPT(1,\a,\b)<0$ & $\dn(\Delta\a,\Delta\B)=\dn(1,\a,\B)<0$ & (\ref{x7})--(\ref{z30}) \\[5pt]
Simon~\cite{Simon}                   & $\dPT(1,\a,\A,\b,\B)<0$ & $\dn(1,\a,\A,\B,\b) +\dn(1,\a,\B)$ & (\ref{x43}) \\[5pt]
 & & ~~~$+\dn(1,\a,\A,\B)+\dn(1,\a,\B,\b)<0$ & \\[5pt]
Mancini {\em et al.}~\cite{Mancini}   & $\dPT(1,\a+\b,\A +\B)<0$ & $\dn(1,\a+\B,\A+\b) + 2 \dn(1,\a+\B) +1<0$ & (\ref{x81}),~(\ref{x57}) \\[5pt]
Hillery \& Zubairy~\cite{Hillery06} & $\dPT(1,\a\b)<0$ & $\dn(1,\a\B)<0$ & (\ref{x1}),~(\ref{x2}) \\[5pt]
{\em ditto} & $\dPT(1,\a^m \b^n)<0$ & $\dn(1,\a^m (\B)^n)<0$ & (\ref{x60})--(\ref{x63}) \\[5pt]
{\em ditto} & $\dPT(\a,\b)<0$ & $\dn(\a,\B)<0$ & (\ref{x4}),~(\ref{x6}) \\[5pt]
{\em ditto} & $\dPT(1,\a\b\c)<0$ & $\dn(1,\A\b\c)<0$ & (\ref{x34}),~(\ref{x46}) \\[5pt]
Miranowicz {\em et al.}~\cite{MPHH09}& $\dPT(\a,\b\c)<0$ & $\dn(\A,\b\c)<0$ & (\ref{x49}) \\[4pt]
Other entanglement tests
 & $\dPT(1,\a^{k} \b^{l}\c^{m})<0$ & $\dn(1,(\A)^{k} \b^{l}\c^{m})<0$ & (\ref{z24}),~(\ref{z25}) \\[5pt]
 & $\dPT(\a^{k}, \b^{l}\c^{m})<0$ & $\dn((\A)^{k}, \b^{l}\c^{m})<0$ & (\ref{z26}),~(\ref{z27}) \\[5pt]
 & $\dPT(1,\a\b,\A \B)<0$ & $\dn(1,\a\B,\A\b) + (\mean{\hat n_{1}+\hat n_{2}}+1)\, \dn(1,\a\B)<0$ & (\ref{x69}),~(\ref{x56}) \\[5pt]
 & $\dPT(1,\a\B,\A\b)<0$ & $\dn(1,\a\b,\A\B)+\na\nb + \mean{\hat n_{1}+\hat n_{2}}\, \dn(1,\a\b)<0$ & (\ref{x75}),~(\ref{x59}) \\[5pt]
 & $\dPT(1,\a+\b,\A +\B)<0$ & $\dn(1,\a+\B,\A+\b) + 2 \dn(1,\a+\B)<0$ & (\ref{x87}),~(\ref{x58}) \\[2pt]
\hline \hline
\end{tabular}
\end{center}
\end{table*}

%------------------------------------------------------------------
\subsection{Nonclassicality and
the Cauchy-Schwarz inequality}

The Cauchy-Schwarz inequality (CSI) for operators can be written
as follows (see, e.g., Ref.~\cite{MandelBook}):
\begin{eqnarray}
  \mean{\hat A^{\dagger} \hat A} \mean{\hat B^{\dagger} \hat B} &\ge&
  |\mean{\hat A^{\dagger} \hat B}|^2,
\label{x92}
\end{eqnarray}
where $\hat A$ and $\hat B$ are arbitrary operators for which the
above expectations exist. Indeed, $\mean{\hat A^{\dagger} \hat
B}\equiv \tr(\rho\hat{A}^\dagger \hat B)$ is a valid inner product
because of the positivity of $\rho$. Similarly, one can define a
valid scalar product for a positive $P$~function. In detail, by
identifying $\hat A=f_1({\bf a,a}^\dagger)$ and $\hat B=f_2({\bf
a,a}^\dagger)$, one can define the scalar product
\begin{equation}
\label{x97}
\begin{split}
\normal{\f_i^{\dagger} \f_j}
    = \intda
f^*_i(\bm{\alpha,\alpha}^*) f_j(\bm{\alpha,\alpha}^*)
  P(\bm{\alpha,\alpha}^*).
\end{split}
\end{equation}
Then, a CSI can be written as:
\begin{eqnarray}
  \normal{\f_1^{\dagger} \f_1}
  \normal{\f_2^{\dagger} \f_2} \cl
  |\normal{\f_1^{\dagger} \f_2}|^2.
\label{x94}
\end{eqnarray}
Such CSI, for a given choice of operators $\f_1$ and $\f_2$, can
be violated by some nonclassical fields described by a
$P$~function which is not positive everywhere, that is such that
\eqref{x97} does not actually define a scalar product. We then say
that the state of the fields violates the CSI. The nonclassicality
of states violating the CSI can be shown by analyzing Criterion~3
for $\F=(\f_1,\f_2)$, which results in
\begin{eqnarray}
  \dfn &=& \DET{\normal{\f_1^{\dagger} \f_1}
  & \normal{\f_1^{\dagger} \f_2}}{
  \normal{\f_1 \f_2^{\dagger}}
  &\normal{\f_2^{\dagger} \f_2}} \ncl 0.
\label{x96}
\end{eqnarray}

%------------------------------------------------------------------
\subsection{A zoo of nonclassical phenomena\label{Sect2b}}

In Table I, we present a variety of multimode nonclassicality
criteria, which can be derived by applying Criterion 3 as shown in
this subsection and in greater detail in Appendices A--C.

In the following, we give a few simple examples of other classical
inequalities, which---to our knowledge---have not been discussed
in the literature. In particular, we analyze inequalities based on
determinants of the following form:
\begin{eqnarray}
  D(x,y,z) &=& \left|
  \begin{array}{lll}
    1 & x & x^* \\
    x^* & z & y^* \\
    x & y & z \
  \end{array}
  \right|.
\label{x66}
\end{eqnarray}
%------------------------------------------------------------------
%%% I
(i) By applying Criterion~3 for $\F=(1,\a\b,\A\B)$, we obtain
\begin{equation}
  \dfn=D(\<\a\b\>,\<\a^2\b^2\>,\<\hat n_1\hat n_2\>)\ncl 0,
\label{x72}
\end{equation}
where $\hat n_1=\A \a$ and $\hat n_2=\B \b$.

%------------------------------------------------------------------
%%% II
\noindent (ii) For $\F=(1,\a\B,\A \b)$ one obtains
%%% 2'
\begin{equation}
  \dfn=D(\<\a\B\>,\<\a^2(\B)^2\>,\<\hat n_1\hat n_2\>)\ncl 0.
\label{x78}
\end{equation}
(iii) For $\F=(1,\a+\B,\A +\b)$, Criterion~3 leads to
%------------------------------------------------------------------
%%% III
\begin{equation}
  \dfn=D(\<\a+\B\>,\<(\a+\B)^2\>,z)\ncl 0,
\label{x84}
\end{equation}
where $z=\<\hat n_1\>+\<\hat n_2\>+2{\rm Re}\<\a\b\>$.

%------------------------------------------------------------------
%%% IV
\noindent (iv) For $\F=(1,\a+\b,\A +\B)$ one has
\begin{equation}
  \dfn=D(\<\a+\b\>,\<(\a+\b)^2\>,z)\ncl 0,
\label{x90}
\end{equation}
where  $z=\<\hat n_1\>+\<\hat n_2\>+2{\rm Re}\<\a\B\>$. These
nonclassicality criteria, given by Eqs.~(\ref{x72})--(\ref{x90}),
will be related to the entanglement criteria in
subsection~\ref{Sect3c2}.

Another example, which is closely related to the Simon
entanglement criterion~\cite{Simon}, as will be shown in
subsection~\ref{Sect3c1}, can be obtained from Criterion~3
assuming $\F=(1,\a,\A,\B,\b)$. Thus, we obtain:
\begin{equation}
  \dfn= \left|
  \begin{array}{lllll}
    1       & \mean{\a} & \mean{\A} & \mean{\B} & \mean{\b} \\
    \mean{\A} & \mean{\A\a} & \mean{(\A)^2} & \mean{\A\B} & \mean{\A\b} \\
    \mean{\a} & \mean{\a^2} & \mean{\A\a} & \mean{\a\B} & \mean{\a\b} \\
    \mean{\b} & \mean{\a\b} & \mean{\A\b} & \mean{\B\b} & \mean{\b^2} \\
    \mean{\B} & \mean{\a\B} & \mean{\A\B} & \mean{(\B)^2} & \mean{\B\b} \\
  \end{array}
  \right| \ncl 0.
\label{x36}
\end{equation}

%------------------------------------------------------------------
\section{Entanglement and nonclassicality criteria}

Here, we express various two- and three-mode entanglement
inequalities in terms of nonclassicality inequalities derived from
Criterion~3, which are summarized in Table~II. First, we briefly
describe the Shchukin-Vogel entanglement criterion, which enables
the derivation of various entanglement inequalities.

%------------------------------------------------------------------
\subsection{The Shchukin-Vogel entanglement criterion\label{Sect3a}}

The Criterion~3 of nonclassicality resembles the Shchukin-Vogel
(SV) criterion~\cite{SV05,MP06,MPHH09} for distinguishing states
with positive partial transposition (PPT) from those with
nonpositive partial transposition (NPT). In analogy to
Eqs.~(\ref{N06}) and~(\ref{N07}), one can define a matrix
$M(\hat\rho)=[M_{ij}(\hat\rho)]$ of moments as follows:
\begin{equation}
M_{ij}(\hat\rho) =\tr\big[({\hat a}^{\dagger i_{1}}{\hat
a}^{i_{2}}{\hat b}^{\dagger i_{3}}{\hat b}^{i_{4}})^\dagger ({\hat
a}^{\dagger j_{1}}{\hat a}^{j_{2}}{\hat b} ^{\dagger j_{3}}{\hat
b}^{j_{4}})\hat\rho\big], \label{N06x}
\end{equation}
where the subscripts $i$ and $j$ correspond to multi-indices
$(i_1,i_2,i_3,i_4)$ and $(j_1,j_2,j_3,j_4)$, respectively. Note
that, contrary to Eq.~(\ref{N06}), the creation and annihilation
operators are {\em not} normally ordered. As discussed in
Ref.~\cite{MPHH09}, the matrix $M(\hat\rho)$ of moments for a
separable state $\hat\rho$ is also separable, i.e.,
\begin{equation}
  \hat\rho=\sum_i p_i
\hat\rho^A_i\otimes\hat\rho^B_i \Rightarrow M(\hat\rho)=\sum_i p_i
M^A(\hat\rho^A_i)\otimes M^B(\hat\rho^A_i),
 \label{Nsep}
\end{equation}
where $p_i\geq0$, $\sum_ip_i=1$, $M^A(\hat\rho^A)=
\sum_{i'j'}M_{i'j'}(\Hat\rho^A)|i'\rangle\langle j'|$ is expressed
in a formal basis $\{|i'\rangle\}$ with $i'=(i_1,i_2,0,0)$ and
$j'=(j_1,j_2,0,0)$; $M^B(\Hat\rho^B)$ defined analogously.
Reference~\cite{SV05} proved the following criterion:
\begin{criterion}
A bipartite quantum state $\hat\rho$ is NPT if and only if
$M(\hat\rho^\Gamma)$ is NPT. \label{t1}
\end{criterion}
The elements of the matrix of moments,
$M(\hat\rho^\Gamma)=[M_{ij}(\hat\rho^\Gamma)]$, where $\Gamma$
denotes partial transposition in some fixed basis, can be simply
calculated as
\begin{eqnarray}
M_{ij}(\hat\rho^\Gamma)
% =M_{i_1i_2i_3i_4,j_1j_2j_3j_4}(\hat\rho^\Gamma)
=\tr\big[({\hat a}^{\dagger i_{1}}{\hat a}^{i_{2}}{\hat
b}^{\dagger i_{3}}{\hat b}^{i_{4}})^\dagger ({\hat a}^{\dagger
j_{1}}{\hat a}^{j_{2}}{\hat b} ^{\dagger j_{3}}{\hat
b}^{j_{4}})\hat\rho^\Gamma\big] \nonumber\\
=\tr\big[({\hat a}^{\dagger i_{1}}{\hat a}^{i_{2}}{\hat
b}^{\dagger j_{3}}{\hat b}^{j_{4}})^\dagger ({\hat a}^{\dagger
j_{1}}{\hat a}^{j_{2}}{\hat b} ^{\dagger i_{3}}{\hat
b}^{i_{4}})\hat\rho\big].\; \label{N06y}
\end{eqnarray}
Let us define
\begin{eqnarray}
 d^{\Gamma}_{\F}(\hat\rho)=
 \Det{
 \langle\f_{r_1}^\dagger \f_{r_1}\rangle^{\Gamma} &
 \langle\f_{r_1}^\dagger \f_{r_2}\rangle^{\Gamma} & \cdots &
 \langle\f_{r_1}^\dagger \f_{r_N} \rangle^{\Gamma}
}{
 \langle\f_{r_2}^\dagger \f_{r_1}\rangle^{\Gamma} &
 \langle\f_{r_2}^\dagger \f_{r_2}\rangle^{\Gamma} & \cdots &
 \langle\f_{r_2}^\dagger \f_{r_N}\rangle^{\Gamma}
}{
 \vdots & \vdots & \ddots & \vdots
}{
 \langle\f_{r_N}^\dagger \f_{r_1}\rangle^{\Gamma} &
 \langle\f_{r_N}^\dagger \f_{r_2}\rangle^{\Gamma} & \cdots &
 \langle\f_{r_N}^\dagger \f_{r_N}\rangle^{\Gamma} },
  \label{t4a}
\end{eqnarray}
in terms of $\langle\f_{r_i}^\dagger
\f_{r_j}\rangle^{\Gamma}\equiv \mean{(\f_{r_i}^\dagger
\f_{r_j})^\Gamma}$ ($i,j=1,...,N$). For example, if $\hat X$ is an
operator acting on two or more modes, and we take partial
transposition with respect to the first mode,
$$
\hat X^\Gamma=(T\otimes{\rm id})(\hat X),
$$
with $T$ the transposition acting on the first mode and ${\rm id}$
the identity operation doing nothing on the remaining modes,
respectively. Then the SV Criterion~4, for brevity referred here
to as the entanglement criterion, can be formulated as
follows~\cite{MPHH09}:
\begin{criterion} %5
A bipartite state $\hat\rho$ is NPT if and only if there exists
${\F}$, such that $d^{\Gamma}_{\F}(\hat\rho)$ is negative.
\end{criterion}
This Criterion~5 can be written more compactly as follows:
\begin{eqnarray}
\hat \rho \textrm{~is PPT} &\Leftrightarrow& \forall {\F}: \quad
d^\Gamma_{\F}(\hat\rho) \ge 0,
\nonumber \\
\hat \rho \textrm{~is NPT} &\Leftrightarrow& \exists {\F}: \quad
d^\Gamma_{\F}(\hat\rho) <0. \label{N08b}
\end{eqnarray}
As for the case of the nonclassicality criteria, the original SV
criterion actually refers to a set $\F$ given by monomials in the
creation and annihilation operators. This entanglement criterion
can be applied not only to two-mode fields but also to multimode
fields~\cite{SV06multi,MPHH09}. Note that Criterion~5 does not
detect PPT-entangled states (which are part, and possibly the only
members, of the family of the so-called bound entangled
states)~\cite{HorodeckiReview}. Analogously to the notation of
$\ncl$, we use the symbol $\ent$ to indicate that a given
inequality can be fulfilled {\em only} for entangled states.

Here we show that various well-known entanglement inequalities can
be derived from the nonclassicality Criterion~3 including the
criteria of Hillery and Zubairy \cite{Hillery06}, Duan {\em et
al.} \cite{Duan}, Simon \cite{Simon}, or Mancini {\em et al.}
\cite{Mancini}. We also derive new entanglement criteria and show
their relation to the nonclassicality criterion.

Other examples of entanglement inequalities, which can be easily
derived from nonclassicality criteria,
include~\cite{Raymer,Agarwal05,Song}. However, for brevity, we do
not include them here.

%------------------------------------------------------------------
\subsection{Entanglement and the Cauchy-Schwarz inequality \label{Marco}}

The matrix $M_{\F}^{(n)}(\hat \rho)$ is linear in its state $\hat
\rho=\sum_ip_i\hat \rho_i$. Therefore we have
\begin{eqnarray}
M_{\F}^{(n)}(\hat \rho)=\sum_ip_i M_{\F}^{(n)}(\hat \rho_i)\geq 0
\end{eqnarray}
if $M_{\F}^{(n)}(\hat \rho_i)\geq 0$ for all $\hat \rho_i$. Thus,
$M_{\F}^{(n)}$ is positive for separable states if it is positive
on factorized states.

Let
\begin{eqnarray}
\F=(\f_1,\ldots,\f_N)
\end{eqnarray}
with functions $\f_{i}=\f_{i1}\f_{i2}\cdots \f_{iM}$, where
\begin{eqnarray}
\f_{ij}=\begin{cases}
1 & \textrm{if }i\neq k_j\\
\textrm{either }g_j(\hat a_j)\textrm{ or }g_j(\hat a^\dagger_j)&
\textrm{if }i=k_j.
\end{cases}
\end{eqnarray}
Here, $i$ is the index of the element $\f_i$ in $\F$, and index
$j$ refers to the mode. $\f_{ij}$ is possibly different from the
identity for one unique value $i=k_j$, and in that case it is
equal to a function $g_j$ of either the creation or annihilation
operators of mode $j$, but not of both.

Writing the matrix $M_{\F}^{(n)}$ in a formal basis
$\{|k\rangle\}$, one then has
\begin{eqnarray}
\begin{aligned}
M_{\F}^{(n)}&=\sum_{kl}\langle :\f_k^{\dagger}\f_l : \rangle |k\rangle\langle l |\\
    &=\sum_{kl}\langle :\f_{k1}^{\dagger}\f_{l1}\ldots  \f_{kM}^{\dagger}\f_{lM}: \rangle |k\rangle\langle l |.\\
\end{aligned}
\end{eqnarray}

For factorized states holds
\begin{eqnarray}
\begin{aligned}
M_{\F}^{(n)}&=\sum_{kl}\langle :\f_{k1}^{\dagger}\f_{l1}:\rangle \cdots  \langle : \f_{kM}^{\dagger}\f_{lM}: \rangle |k\rangle\langle l |\\
    &=\sum_{k}\langle :\f_{k1}^{\dagger}\f_{k1}:\rangle \cdots  \langle : \f_{kM}^{\dagger}\f_{kM}: \rangle |k\rangle\langle l |\\
    &\quad+\sum_{k\neq l}\langle :\f_{k1}^{\dagger}\f_{l1}:\rangle \cdots  \langle : \f_{kM}^{\dagger}\f_{lM}: \rangle |k\rangle\langle l |\\
    &=\sum_{k}\langle :\f_{k1}^{\dagger}\f_{k1}:\rangle \cdots  \langle : \f_{kM}^{\dagger}\f_{kM}: \rangle |k\rangle\langle l |\\
    &\quad+\sum_{k\neq l}\langle \f_{k1}^{\dagger}\rangle\langle \f_{l1} \rangle \cdots  \langle \f_{kM}^{\dagger}\rangle\langle \f_{lM}\rangle |k\rangle\langle l |\\
    &\geq \sum_{k}\langle \f_{k1}^{\dagger}\rangle \langle \f_{k1} \rangle \cdots  \langle \f_{kM}^{\dagger}\rangle \langle \f_{kM} \rangle |k\rangle\langle l |\\
    &\quad+\sum_{k\neq l}\langle \f_{k1}^{\dagger}\rangle\langle \f_{l1} \rangle \cdots  \langle \f_{kM}^{\dagger}\rangle\langle \f_{lM}\rangle |k\rangle\langle l |\\
    &=\Big(\sum_{k}\langle \f_{k1}^{\dagger}\rangle \cdots \langle \f_{kM}^{\dagger}\rangle
    |k\rangle\Big) \\
    &\quad\times \Big(\sum_{l}\langle \f_{l1}\rangle \cdots \langle \f_{lM}\rangle \langle l|\Big)\geq0.
\label{Marco1}
\end{aligned}
\end{eqnarray}
The first equality comes from the state being factorized. The
third equality is due to the fact that the $\f_{ij}$s are
functions of either annihilation or creation operators, but not of
both, so  $\langle :\f_{k1}^{\dagger}\f_{l1}:\rangle= \langle
\f_{k1}^{\dagger}\f_{l1}\rangle$ or $\langle
:\f_{k1}^{\dagger}\f_{l1}:\rangle= \langle \f_{l1}
\f_{k1}^{\dagger}\rangle$, and that for $k\neq l$ at least one
among $\f_{k1}^{\dagger}$ and $\f_{l1}$, let us say, e.g.,
$\f_{l1}$, is equal to the identity---in particular this implies
that its expectation value is equal to $\langle \f_{l1}\rangle=1$.
The first inequality is due to the fact that $\langle
:\f_{k1}^{\dagger}\f_{k1}:\rangle= \langle
\f_{k1}^{\dagger}\f_{k1}\rangle$ or ${\langle
:\f_{k1}^{\dagger}\f_{k1}:\rangle}= \langle \f_{k1}
\f_{k1}^{\dagger}\rangle$, and to the Cauchy-Schwarz inequality.

%------------------------------------------------------------------
\subsection{Entanglement criteria {\em equal} to
nonclassicality criteria\label{Sect3b}}

By applying the nonclassicality Criterion~3, we give a few
examples of classical inequalities, which can be violated {\em
only} by entangled states.

%------------------------------------------------------------------
\subsubsection{Hillery-Zubairy's entanglement criteria}

Hillery and Zubairy~\cite{Hillery06} derived a few entanglement
inequalities both for two-mode fields:
\begin{eqnarray}
  \mean{\hat n_1\hat n_2} \ent |\mean{\a\B}|^2,
\label{x1}
  \\
  \mean{\hat n_1}\mean{\hat n_2} \ent |\mean{\a\b}|^2 ,
\label{x4}
\end{eqnarray}
and three-mode fields
\begin{eqnarray}
 \mean{\hat n_1\hat n_2\hat n_3} \ent |\mean{\A\b\c}|^2.
 \label{x34}
\end{eqnarray}
These inequalities can be derived from the entanglement
Criterion~5~\cite{SV05,MPHH09} assuming:  $\F=(1,\a\b)$ to derive
Eq.~(\ref{x1}), $\F=(\a,\b)$ for Eq.~(\ref{x4}), and
$\F=(1,\a\b\c)$ for Eq.~(\ref{x34}).

On the other hand, Eq.~(\ref{x1}) can be obtained from the
nonclassicality Criterion~3 assuming $\F=(1,\a\B)$, which gives
\begin{eqnarray}
  \dfn &=& \DET{1&\mean{\a\B}}{\mean{\A\b}& \mean{\hat n_1\hat n_2}} \ncl 0.
\label{x2}
\end{eqnarray}
Analogously, assuming $\F=(\a,\B)$, one gets
\begin{eqnarray}
  \dfn &=& \DET{\mean{\hat n_1}&\mean{\A\B}}
               {\mean{\a\b}&\mean{\hat n_2}} \ncl 0,
\label{x6}
\end{eqnarray}
which corresponds to Eq.~(\ref{x4}). By choosing a set of
three-mode operators $\F=(1,\A\b\c)$, one readily obtains
\begin{eqnarray}
    \dfn &=& \DET
    {1&\mean{\A\b\c}}
    {\mean{\a\B\C}& \mean{\hat n_1\hat n_2\hat n_3}} \ncl 0, \label{x46}
\end{eqnarray}
which corresponds to Eq.~(\ref{x34}).

By applying Criterion~3 with $\F=(\A,\b\hat a_3)$, we find another
inequality
\begin{eqnarray}
     \dfn &=& \DET
    {\mean{\hat n_1}&\mean{\a\b\c}}
    {\mean{\a\b\c}^*& \mean{\hat n_2\hat n_3}} \ncl 0,
\label{x49}
\end{eqnarray}
which was derived in Ref.~\cite{MPHH09} from the entanglement
Criterion~5.

Using the Cauchy-Schwarz inequality, Hillery and
Zubairy~\cite{Hillery06} also found a more general form of
inequality than the one in Eq.~(\ref{x1}), which reads as follows:
\begin{eqnarray}
  \mean{(\A)^m\a^m (\B)^n\b^n} \ent |\mean{\a^m (\B)^n}|^2.
\label{x60}
\end{eqnarray}
This inequality can be derived from the nonclassicality
Criterion~3 for $\F=(1,\a^m (\B)^n)$, which leads to
\begin{eqnarray}
     \dfn = \DET
    {1&\mean{\a^m (\B)^n}}
    {\mean{(\A)^m \b^n} & \mean{ (\A)^m\a^m (\B)^n\b^n}} \ncl 0. \label{x62}
\end{eqnarray}
Alternatively, Eq.~(\ref{x60}) can be derived from the
entanglement Criterion~5 for $\F=(1,\a^m \b^n)$. Thus, we see that
\begin{equation}
  \dn(1,\a^m (\B)^n) = \dPT(1,\a^m \b^n) \ent 0, \label{x63}
\end{equation}
where, for clarity, we use the notation $d^{k}(\F)$ instead of
$d^{k}_{\F}$ for $k=(n),\Gamma$. Moreover, we can generalize
entanglement inequality, given by Eq.~(\ref{x46}), as follows:
\begin{equation}
\mean{\hat n_1^{k}\hat n_2^{l}\hat n_3^{m} } \ent |\mean{(\A)^{k}
\b^{l}\c^{m}}|^2 \label{z24}
\end{equation}
for arbitrary integers $k,l,m>0$. This inequality can be proved by
applying both Criteria~3 and~5:
\begin{eqnarray}
\dn(1,(\A)^{k} \b^{l}\c^{m})= \dPT(1,\a^{k} \b^{l}\c^{m})
\hspace{3cm}
\nonumber \\
=\DET{1&\mean{(\A)^{k} \b^{l}\c^{m}}} {\<(\a^\dagger)^{k}
\b^{l}\c^{m}\>^*& \mean{\hat n_1^{k}\hat n_2^{l}\hat n_3^{m}
}}\ncl 0, \hspace{1cm} \label{z25}
\end{eqnarray}
where the first mode is partially-transposed. Analogously,
Eq.~(\ref{x49}) can be generalized to following entanglement
inequality:
\begin{equation}
\mean{\hat n_1^{k}}\mean{\hat n_2^{l}\hat n_3^{m} } \ent
|\mean{\a^{k} \b^{l}\c^{m}}|^2, \label{z26}
\end{equation}
which can be shown by applying Criteria~3 and~5:
\begin{eqnarray}
\dn((\A)^{k}, \b^{l}\c^{m})&=& \dPT(\a^{k}, \b^{l}\c^{m})
\nonumber \\
&=&\DET{\mean{\hat n_1^{k}}&\mean{\a^{k} \b^{l}\c^{m}}} {\<\a^{k}
\b^{l}\c^{m}\>^* & \mean{\hat n_2^{l}\hat n_3^{m} }} \ncl 0.
\hspace{5mm} \label{z27}
\end{eqnarray}

It is worth remarking that in all the above cases, once the $\ncl$
inequalities are found as nonclassicality inequalities, it is easy
to check that they can be satisfied only by entangled states, that
is they really are $\ent$ inequalities. Indeed, the determinant
condition is the only nontrivial one for establishing the
positivity of the involved $2\times 2$ matrices. Further, these
matrices are linear in the state with respect to which the
expectation values are calculated. Thus, if we prove that the
matrices are positive for factorized states, then we have that
they are necessarily positive for a separable state, and so are
the determinants. For the sake of concreteness and clarity, we
prove the positivity of the $2\times 2$ matrix of Eq.~\eqref{x6}
for a factorized state. The positivity of the other matrices for
factorized states is analogously proved.

For a factorized state, as a special case of inequalities given in
Eq.~(\ref{Marco1}), we have
\begin{equation}
\begin{aligned}
\left(
\begin{array}{cc}
\mean{\hat n_1}&\mean{\A\B}\\
               \mean{\a\b}&\mean{\hat n_2}
\end{array}
\right) &= \left(
\begin{array}{cc}
\mean{\A\a}&\mean{\A}\mean{\B}\\
               \mean{\a}\mean{\b}&\mean{\B\b}
\end{array}
\right)\\
&\geq \left(
\begin{array}{cc}
\mean{\A}\mean{\a}&\mean{\A}\mean{\B}\\
               \mean{\a}\mean{\b}&\mean{\B}\mean{\b}
\end{array}
\right)\\
&= \left(
\begin{array}{c}
\mean{\A}\\
\mean{\b}
\end{array}
\right) \left(
\begin{array}{cc}
\mean{\a}&\mean{\B}
\end{array}
\right)\geq0,
\end{aligned}
\end{equation}
where the first inequality is due to the Cauchy-Schwarz inequality
$\mean{\hat{X}^\dagger\hat{X}}\geq|\mean{\hat{X}}|^2$.

%------------------------------------------------------------------
\subsubsection{Entanglement criterion of Duan et al.}

A sharpened version of the entanglement criterion of Duan {\em et
al.}~\cite{Duan} can be formulated as follows~\cite{SV05}:
\begin{eqnarray}
   \mean{\Delta\A\Delta\a}\mean{\Delta\B\Delta\b}\ent
   |\mean{\Delta\a\Delta\b}|^2,
\label{x7}
\end{eqnarray}
where $\Delta\hat a_i=\hat a_i-\mean{\hat a_i}$ for $i=1,2$.
Equation~(\ref{x7}) follows from the entanglement Criterion~5 for
$\F=(1,\a,\b)$~\cite{SV05} or, equivalently, for
$\F=(\Delta\a,\Delta\b)$. It can also be derived from the
nonclassicality Criterion~3 for $\F=(\Delta\a,\Delta\B)$. Thus, we
obtain
\begin{eqnarray}
\dfn &=& \DET
    {\mean{\Delta\A\Delta\a}&\mean{\Delta\A\Delta\B}}
    {\mean{\Delta\a\Delta\b}& \mean{\Delta\B\Delta\b}} \ncl 0,
\label{x9}
\end{eqnarray}
which corresponds to Eq.~(\ref{x7}). Alternatively, by choosing
$\hat F=(1,\a,\B)$, one obtains
\begin{equation}
\dfn = \DETT {1&\mean{\a}&\mean{\B}} {\mean{\A}&\na&\mean{\A\B}}
{\mean{\b}&\mean{\a\b}&\nb}, \label{z30}
\end{equation}
which is equal to Eq.~(\ref{x9}). Thus, it is seen that this
nonclassicality criterion is equal to the entanglement
criterion. Moreover, the advantage of using polynomials, instead
of monomial, functions of moments in $\F$ is apparent. The same
conclusion was drawn by comparing Eqs.~(\ref{N18}) and~(\ref{z36})
or Eqs.~(\ref{x27}) and~(\ref{z34}).

%------------------------------------------------------------------
\subsection{Entanglement criteria via sums of nonclassicality
criteria\label{Sect3c}}

Here, we present a few examples of classical inequalities derived
from the entanglement Criterion~5 and the nonclassicality
Criterion~3 that are apparently not equal. More specifically, we
have presented in subsection~\ref{Sect3b} examples of classical
inequalities, which can be derived from the entanglement
Criterion~5 for a given $\F_1$ or, equivalently, from the
nonclassicality Criterion~3 for $\F_2$ equal to a partial
transpose of $\F_1$. In this section, we give examples of
entanglement inequalities, which {\em cannot} be derived from
Criterion~3 for $\F_2=\F_1^\Gamma$.

States satisfying Criterion~5 for entanglement must be
nonclassical, as any entangled state is necessary nonclassical in
the sense of Criterion 1. We will provide specific examples that
satisfying an entanglement inequality implies satisfying one or
more nonclassical inequalities. This approach enables an analysis
of the entanglement for a given nonclassicality. The main problem
is to express $\dfPT\equiv\dPT(\F)$ as linear combinations of some
$\dn(\F^{(k)})$, i.e.:
\begin{eqnarray}
  \dfPT = \sum_k c_k \dn(\F^{(k)}),
\label{x52}
\end{eqnarray}
where $c_k>0$. To find such expansions explicitly, we apply the
following three properties of determinants: (i) The Laplace
expansion formula along any row (or column): $\det
M=\sum_{j}(-1)^{i+j}M_{ij}\mu_{ij}$, where $\mu_{ij}$ is a minor
of a matrix $M=(M_{ij})$. (ii) Swapping rule: By exchanging any
two rows (columns) of a determinant, the value of the determinant
is the same of the original determinant but with opposite sign.
(iii) Summation rule: If some (or all) the elements of a column
(row) are sum of two terms, then the determinant can be given as
the sum of two determinants, e.g.,
$\det(a+a',b+b';c,d)=\det(a,b;c,d)+\det(a',b';c,d).$

%------------------------------------------------------------------
\subsubsection{Simon's entanglement criterion\label{Sect3c1}}

As the first example of such nontrivial relation between the
nonclassicality and entanglement criteria, let us consider Simon's
entanglement criterion~\cite{Simon}. As shown in Ref.~\cite{SV05},
it can be obtained from Criterion~5 as $\dfPT\ent 0$ for
$\F=(1,\a,\A,\b,\B)$. We found that Simon's criterion can be
expressed as a sum of nonclassicality criteria as follows:
\begin{eqnarray}
  \dfPT &=&
  \dn(1,\a,\A,\B,\b) +\dn(1,\a,\B)
\nonumber\\
  &&+\dn(1,\a,\A,\B)+\dn(1,\a,\B,\b),
\label{x43}
\end{eqnarray}
where $\dn(1,\a,\A,\B,\b)$ is given by Eq.~(\ref{x36}). Moreover,
$\dfn$ for $\F=(1,\a,\A,\B)$, $\F=(1,\a,\B,\b)$ and $\F=(1,\a,\B)$
can be obtained from~(\ref{x36}) by analyzing its principal
minors. Thus, one can prove the entanglement for a given
nonclassicality by checking the violation of specific classical
inequalities resulting from the nonclassicality Criterion~3.

%------------------------------------------------------------------
\subsubsection{Other entanglement criteria\label{Sect3c2}}

Now, we present a few entanglement inequalities, which are simpler
than Simon's criterion, but still correspond to sums of
nonclassicality inequalities.

Let us denote the following determinant:
\begin{eqnarray}
  D(x,y,z,z') &=& \left|
  \begin{array}{lll}
    1 & x & x^* \\
    x^* & z & y^* \\
    x & y & z' \
  \end{array}
  \right|.
\label{x65}
\end{eqnarray}
%------------------------------------------------------------------
%%% I
(i) Criterion~5 for $\F=(1,\a\b,\A \B)$ results in
\begin{equation}
  \dfPT=D\left(\<\a\B\>,\<\a^2(\B)^2\>,
  \mn,z'\right)\ent 0,
\label{x69}
\end{equation}
where $z'=\mean{(\hat n_1+1)(\hat n_2+1)}$. By using the
aforementioned properties of determinants, we find that the
entanglement criterion in Eq.~(\ref{x69}) can be given as the
following sum of nonclassicality inequalities resulting from
Criterion~3:
\begin{eqnarray}
  \dfPT &=& \dn(1,\a\B,\A\b)
\nonumber \\
  &&+ (\na +\nb+1)\, \dn(1,\a\B). \label{x56}
\end{eqnarray}
%------------------------------------------------------------------
%%% II
(ii) Criterion~5 for $\F=(1,\a\B,\A\b)$ leads to
\begin{equation}
  \dfPT=D(\<\a\b\>,\<\a^2\b^2\>,z,z')\ent 0,
\label{x75}
\end{equation}
where $z=\mn +\na $ and $z'=\mn +\nb$. Analogously to
Eq.~(\ref{x56}), we find that the following sum of the
nonclassicality criteria corresponds to the entanglement criterion
in Eq.~(\ref{x75}):
\begin{eqnarray}
  \dfPT &=& \dn(1,\a\b,\A\B)+\na\nb
\nonumber \\
  &&+ (\na +\nb)\, \dn(1,\a\b). \label{x59}
\end{eqnarray}
%------------------------------------------------------------------
%%% III
(iii) For $\F=(1,\a+\B,\A +\b)$, one obtains
\begin{equation}
  \dfPT=D(\<\a+\b\>,\<(\a+\b)^2\>,z,z)\ent 0,
\label{x81}
\end{equation}
where  $z=\<\hat n_1\>+\<\hat n_2\>+2{\rm Re}\<\a\B\>+1$.
Analogously to the former cases, we find the relation between the
entanglement criterion in Eq.~(\ref{x81}) and the nonclassicality
Criterion~3 as follows:
\begin{eqnarray}
  \dfPT &=& \dn(1,\a+\b,\A+\B)
\notag\\
  &&
  + 2 \dn(1,\a+\b) +1. \label{x57}
\end{eqnarray}
%------------------------------------------------------------------
%%% IV
(iv) As a final example, let us consider the entanglement
Criterion~5 for $\F=(1,\a+\b,\A +\B)$. One obtains
\begin{equation}
  \dfPT=D(\<\a+\B\>,\<(\a+\B)^2\>,z,z')\ent 0,
\label{x87}
\end{equation}
where  $z=\<\hat n_1\>+\<\hat n_2\>+2{\rm Re}\<\a\b\>$ and
$z'=z+2$, which is related to the nonclassicality Criterion~3 as
follows:
\begin{equation}
  \dfPT = \dn(1,\a+\B,\A+\b) + 2 \dn(1,\a+\B), \label{x58}
\end{equation}
where $\dn(1,\a+\B,\A+\b)$ is given by Eq.~(\ref{x84}), and
$\dn(1,\a+\B)$ is given by its principal minor.
Equation~(\ref{x87}) corresponds to the entanglement criterion of
Mancini {\em et al.} \cite{Mancini} (see also~\cite{SV05}).

%------------------------------------------------------------------
\section{Conclusions\label{Sect4}}

We derived classical inequalities for multimode bosonic fields,
which can {\em only} be violated by {\em nonclassical} fields, so
they can serve as a nonclassicality (or quantumness) test. Our
criteria are based on Vogel's criterion~\cite{Vogel08}, which is a
generalization of analogous criteria for single-mode fields of
Agarwal and Tara~\cite{Agarwal92} and, more directly, of Shchukin,
Richter, and Vogel (SRV)~\cite{NCL1,NCL2}. The nonclassicality
criteria correspond to analyzing the positivity of matrices of
normally ordered moments of, e.g., annihilation and creation
operators,  which, by virtue of Sylvester's criterion, correspond
to analyzing the positivity of Glauber-Sudarshan $P$~function. We
used not only monomial, but also polynomial functions of moments.
We showed that this approach can enable simpler and more intuitive
derivation of physically relevant inequalities.

We demonstrated how the nonclassicality criteria introduced here
easily reduce to the well-known inequalities~(see, e.g.,
textbooks~\cite{DodonovBook,VogelBook,MandelBook,PerinaBook},
reviews~\cite{Walls79,Loudon80,Loudon87,Klyshko96}, and
Refs.~\cite{Yuen76,Kozierowski77,Caves85,Reid86,Dalton86,Schleich87,Agarwal88,Luks88,Hillery89,Lee90,Zou90,Klyshko96pla,Miranowicz99a,Miranowicz99b,An99,An00,Jakob01})
describing various multimode nonclassical effects, for short
referred to as the nonclassicality inequalities. Our examples,
summarized in Tables~I and~II, include the following:

(i)~Multimode quadrature squeezing~\cite{VogelBook} and its
generalizations, including the sum and difference squeezing
defined by Hillery~\cite{Hillery89}, and An and
Tinh~\cite{An99,An00}, as well the principal squeezing related to
the Schr\"odinger-Robertson indeterminacy relation~\cite{SR} as
defined by Luk\v{s} {\em et al.}~\cite{Luks88}.

(ii)~Single-time photon-number correlations of two modes,
including squeezing of the sum and difference of photon numbers
(which is also referred to as the photon-number sum/difference
sub-Poisson photon-number statistics)~\cite{PerinaBook},
violations of the Cauchy-Schwarz inequality~\cite{MandelBook} and
violations of the Muirhead inequality~\cite{Muirhead,Lee90}, which
is a generalization of the arithmetic-geometric mean inequality.

(iii)~Two-time photon-number correlations of single modes
including photon
antibunching~\cite{VogelBook,MandelBook,Miranowicz99a} and photon
hyperbunching~\cite{Jakob01,Miranowicz99b} for stationary and
nonstationary fields.

(iv)~Two- and three-mode quantum entanglement inequalities (e.g.,
Refs.~\cite{Duan,Hillery06,Simon,Mancini}). We have shown that
some known entanglement inequalities (e.g., of Duan {\em et
al.}~\cite{Duan}, and Hillery and Zubairy~\cite{Hillery06}) can be
derived as nonclassical inequalities. Other entanglement
inequalities (e.g., of Simon~\cite{Simon}) can be represented by
sums of nonclassicality inequalities.

Moreover, we developed a general method of expressing inequalities
derived from the Shchukin-Vogel entanglement
criterion~\cite{SV05,MP06} as a sum of inequalities derived from
the nonclassicality criteria. This approach enables a deeper
analysis of the entanglement for a given nonclassicality.

We also presented a few inequalities derived from the
nonclassicality and entanglement criteria, which to our knowledge
have not yet been described in the literature.

It is seen that the nonclassicality criteria based on matrices of
moments offer an effective way to derive specific inequalities
which might be useful in the verification of nonclassicality of
particular states generated in experiments.

It seems that the quantum-information community more or less
ignores nonclassicality as something closely related to quantum
entanglement. We hope that this article presents a useful approach
in the direction of a common treatment of both types of phenomena.

%------------------------------------------------------------------
\begin{acknowledgments}
We are very grateful to Marco Piani for his help in clarifying and
generalizing some results of this article. We also thank Werner
Vogel and Jan Sperling for their comments. A.M. acknowledges
support from the Polish Ministry of Science and Higher Education
under Grant No. 2619/B/H03/2010/38. X.W. was supported by the
National Natural Science Foundation of China under Grant No.
10874151, the National Fundamental Research Programs of China
under Grant No. 2006CB921205, and Program for New Century
Excellent Talents in University (NCET). Y.X.L. was supported by
the National Natural Science Foundation of China under Grant No.
10975080. F.N. acknowledges partial support from the National
Security Agency, Laboratory of Physical Sciences, Army Research
Office, National Science Foundation Grant No. 0726909, JSPS-RFBR
Contract No. 09-02-92114, Grant-in-Aid for Scientific Research
(S), MEXT Kakenhi on Quantum Cybernetics, and FIRST (Funding
Program for Innovative R\&D on S\&T).
\end{acknowledgments}

\begin{appendix}

%------------------------------------------------------------------
\section{Unified derivations of criteria for quadrature squeezing
and its generalizations}

Here and in the following appendices, we present a unified
derivation of the known criteria for various multimode
nonclassicality phenomena, which are summarized in Table~I.

%------------------------------------------------------------------
\subsection{Multi-mode quadrature squeezing}

The {\em quadrature squeezing} of multimode fields can be defined
by a negative value of the normally ordered variance
\cite{Caves85,Loudon87,VogelBook}
%------------------------------------------------------------------
\begin{equation}
\varn{X_{\bm{\phi}}}<0 \label{N10}
\end{equation}
with $\Delta\hat{X}_{\bm{\phi}}
=\hat{X}_{\bm{\phi}}-\langle\hat{X}_{\bm{\phi}}\rangle$, of the
multimode quadrature operator
%------------------------------------------------------------------
\begin{equation}
  \hat X_{\bm{\phi}} = \sum_{m=1}^M c_m\; \hat
  x_m(\phi_m),
\label{N11}
\end{equation}
which is given in terms of single-mode phase-rotated quadratures
%------------------------------------------------------------------
\begin{equation}
  \hat x_m(\phi_m)= \hat a_m \exp(i\phi_m)
  + \hat a_m^\dagger \exp(-i\phi_m). \label{N12}
\end{equation}
It is a straightforward generalization of the single-mode
quadrature squeezing~\cite{Yuen76,Walls79}. In~(\ref{N11}),
$\bm{\phi}=(\phi_1,...,\phi_M)$ and $c_m$ are real parameters. In
the analysis of physical systems, it is convenient to analyze the
annihilation ($\hat a_m$) and creation ($\hat a_m^\dagger$)
operators corresponding to slowly-varying operators. Usually,
$\hat x_m(0)$ and $\hat x_m(\pi/2)$ are interpreted as canonical
position and momentum operators, although this interpretation can
be applied for any two quadratures of orthogonal phases, $\hat
x_m(\phi_m)$ and $\hat x_m(\phi_m+\pi/2)$.

The normally ordered variance can be directly calculated from the
$P$~function as follows:
%------------------------------------------------------------------
\begin{equation}
  \varn{X_{\bm{\phi}}} = \intda P(\bm{\alpha,\alpha}^*)[X_{\bm{\phi}}
  (\bm{\alpha,\alpha}^*)-\<\hat X_{\bm{\phi}}\>]^2, \label{N13}
\end{equation}
where
%------------------------------------------------------------------
\begin{equation}
 X_{\bm{\phi}}(\bm{\alpha,\alpha}^*) = \sum_{m=1}^M c_m (\alpha_m e^{i\phi_m} +\alpha^*_m
e^{-i\phi_m}) \label{N14}
\end{equation}
and $\bm{\alpha}=(\alpha_1,...,\alpha_M)$. From Eq.~(\ref{N13}) it
is seen that a negative value of $\varn{X_{\bm{\phi}}}$ implies
the nonpositivity of the $P$~function in some regions of phase
space, so the multimode quadrature squeezing is a nonclassical
effect. This conclusion can also be drawn by applying Criterion~3.
In fact, by choosing $\F=(1,\hat X_{\bm{\phi}})$, one obtains
%------------------------------------------------------------------
\begin{equation}
d_{\F}^{\rm (n)} =
 \DET{1 & \<\hat X_{\bm{\phi}}\>}
 {\<\hat X_{\bm{\phi}}\> & \<:\hat X_{\bm{\phi}}^2:\>}
 = \varn{X_{\bm{\phi}}} \ncl 0,
\label{N15}
\end{equation}
which is the squeezing condition (\ref{N10}).

%------------------------------------------------------------------
\subsection{Two-mode principal squeezing}

For simplicity, we analyze below the two-mode ($M=2$) case for
$c_1=c_2=1$ and $\phi_2-\phi_1=\pi/2$. The two-mode {\em
principal} (quadrature) squeezing can be defined as the
$\bm{\phi}$-optimized squeezing defined by Eq.~(\ref{N10}):
%------------------------------------------------------------------
\begin{equation}
  \min_{\bm{\phi}:\phi_2-\phi_1=\pi/2} \varn{X_{\bm{\phi}}} < 0.
\label{N16}
\end{equation}
By applying the Schr\"odinger-Robertson indeterminacy
relation~\cite{SR}, Luk\v{s} {\em et al.}~\cite{Luks88} have given
the following necessary and sufficient condition for the two-mode
{\em principal squeezing}
%------------------------------------------------------------------
\begin{equation}
% q_{\rm ps} \equiv
\<\Delta \hat a_{12}^\dagger \Delta \hat a_{12} \> <  |\langle
(\Delta \hat a_{12})^2 \rangle|  , \label{N17}
\end{equation}
where $$\hat a_{12}=\hat a_{1}+\hat a_{2},\quad\Delta \hat
a_{12}=\hat a_{12}-\<\hat a_{12}\>.$$ This condition for principal
squeezing can be derived from Criterion~3 by choosing $\F=(\Delta
\hat a_{12}^\dagger,\Delta \hat a_{12})$, which leads to:
%------------------------------------------------------------------
\begin{equation}
  d_{\F}^{\rm (n)} =
 \DET{\<\Delta \hat a_{12}^\dagger \Delta \hat a_{12} \>
 & \langle (\Delta \hat a_{12})^2 \rangle}
 {\langle (\Delta \hat a_{12}^\dagger)^2 \rangle
 & \<\Delta \hat a_{12}^\dagger \Delta \hat a_{12} \>} \ncl 0.
\label{N18}
\end{equation}
Equivalently, by applying Criterion~3 for $\F=(1, \hat
a_{12}^\dagger, \hat a_{12})$ one obtains:
%------------------------------------------------------------------
\begin{equation}
d_{\F}^{\rm (n)} =\DETT{1&\mean{\hat a^\dagger_{12}}&\mean{\hat
a_{12}}} {\mean{\hat a_{12}}&\mean{\hat n_{12}}&\mean{(\hat
a_{12})^2}} {\mean{\hat a^\dagger_{12}}&\mean{(\hat
a^\dagger_{12})^2} &\mean{\hat n_{12}}}, \label{z36}
\end{equation}
where $$\hat n_{12}=\hat a^\dagger_{12}\hat a_{12}=\hat n_1+\hat
n_2 +2{\rm Re}(\A\b).$$ The determinants, given by
Eqs.~(\ref{N18}) and~(\ref{z36}) are equal to each other and
equivalent to Eq.~(\ref{N17}). This example shows that the
application of polynomial functions of moments, instead of
monomials, can lead to matrices of moments of lower dimension.
Thus, the polynomial-based approach can enable simpler and more
intuitive derivations of physically relevant criteria.

%------------------------------------------------------------------
\subsection{Sum squeezing}

According to Hillery~\cite{Hillery89}, a two-mode state exhibits
{\em sum squeezing} in the direction $\phi$ if the variance of
\begin{eqnarray}
  \hat V_{\phi} &=& \frac12 (\hat a_1 \hat a_2 e^{-i\phi}+
  \hat a_1^\dagger \hat a_2^\dagger e^{i\phi} ) \label{N19}
\end{eqnarray}
satisfies
\begin{eqnarray}
   \var{V_{\phi}} &<& \frac12 \<\hat V_z\>,
\label{N20}
\end{eqnarray}
where $$\hat V_z=\frac12(\hat n_1+\hat n_2+1)$$ and $\hat n_m=\hat
a_m^\dagger  \hat a_m$ for $m=1,2$. As for the case of quadrature
squeezing, $\hat a_1$ and $\hat a_2$ usually correspond to slowly
varying operators. Let us denote $\hat V_x=\hat V(\phi=0)$ and
$\hat V_y=\hat V(\phi=\pi/2)$. It is worth mentioning that the
operators $\hat V_x$, $(-\hat V_y)$ and $\hat V_z$ are the
generators of the SU(1,1) Lie algebra. Equation~(\ref{N20}) can be
readily justified by noting that $[\hat V_x,\hat V_y]=i\hat V_z$,
which implies the Heisenberg uncertainty relation
$$\var{V_{x}}\var{V_{y}}\ge \frac14\<\hat V_z\>^2.$$ By analogy with the
standard quadrature squeezing, sum squeezing occurs when
$\min\{\var{V_{x}},\var{V_{y}}\}<\<\hat V_z\>/2$, or more
generally if Eq.~(\ref{N20}) is satisfied. We note that, in
analogy to the principal quadrature squeezing, one can define the
principal sum squeezing by minimizing $\var{V_{\phi}}$ over
$\phi$:
\begin{equation}
  \min_{\phi}\var{V_{\phi}} < \frac12 \<\hat V_z\>.
\label{N20a}
\end{equation}
Conditions~(\ref{N20}) and~(\ref{N20a}) can be easily derived from
Criterion~3. In fact, by noting that
\begin{eqnarray}
  \var{V_{\phi}} &=& \varn{V_{\phi}} + \frac12 \<\hat V_z\>,
  \label{N21}
\end{eqnarray}
the condition for sum squeezing can equivalently be given by a
negative value of the variance $\varn{V_{\phi}}$. On the other
hand, by applying Criterion~3 for $\F=(1,\hat V_{\phi})$, one
obtains
\begin{eqnarray}
  d_{\F}^{\rm (n)} =
  \DET{1 & \mean{\hat V_{\phi}}}
  {\mean{\hat V_{\phi}} & \mean{:\hat V^2_{\phi}:}}
  =\varn{V_{\phi}} \ncl 0,
\label{N22}
\end{eqnarray}
which is equivalent to Eq.~(\ref{N20}). So it is seen that sum
squeezing is a nonclassical effect---in the sense of Criterion 1.

Two-mode sum squeezing can be generalized for any number of modes
by defining the following $M$-mode phase-dependent
operator~\cite{An99}:
\begin{equation}
\hat {\cal V}_\phi = \frac12 \left( {\rm e}^{-i \phi}\prod_{j}
\hat a_j+ {\rm e}^{i \phi}\prod_{j} \hat a_j^\dagger \right)
 \label{z2}
\end{equation}
satisfying the commutation relation
\begin{equation}
[\hat {\cal V}_\phi ,\hat {\cal V}_{\phi+\pi/2} ]=\frac{i}2 \hat
C, \quad\hat C=\prod_{j}(1+\hat n_j)-\prod_{j}\hat n_j. \label{z5}
\end{equation}
Hereafter $j=1,...,M$ and we note that $|\mean{\hat C}|=\mean{\hat
C}$. Thus, multimode sum squeezing along the direction $\phi$
occurs if
\begin{equation}
\var{{\cal V}_\phi} < \frac{|\mean{\hat C}|}4. \label{z9}
\end{equation}
One can find that
\begin{equation}
\var{{\cal V}_\phi}=\varn{{\cal V}_\phi}+ \frac{|\mean{\hat C}|}4.
\label{z6}
\end{equation}
Thus, by applying the nonclassicality Criterion~3 for $\hat
F=(1,\hat {\cal V}_\phi)$, we obtain the sum squeezing condition
\begin{equation}
\varn{{\cal V}_\phi} = \dfn \ncl 0, \label{z10}
\end{equation}
which is equivalent to condition in Eq.~(\ref{z9}).

%------------------------------------------------------------------
\subsection{Difference squeezing}

As defined by Hillery~\cite{Hillery89}, a two-mode state exhibits
{\em difference squeezing} in the direction $\phi$ if
\begin{eqnarray}
 \var{W_{\phi}} &<& \frac12 |\<\hat W_z\>|,
\label{N23}
\end{eqnarray}
where
\begin{eqnarray}
  \hat W_{\phi} &=& \frac12 (\hat a_1 \hat a_2^\dagger  e^{i\phi}+
  \hat a_1^\dagger \hat a_2 e^{-i\phi} ) \label{N24}
\end{eqnarray}
and $\hat W_z=\frac12(\hat n_1-\hat n_2)$. The principal
difference squeezing can be defined as:
\begin{equation}
 \min_{\phi}\var{W_{\phi}} < \frac12 |\<\hat W_z\>|,
\label{N23a}
\end{equation}
in analogy to the principal quadrature squeezing and the principal
sum squeezing. Contrary to the $\hat V_{i}$ operators for sum
squeezing, operators $\hat W_x=\hat W(\phi=0)$, $\hat W_y=\hat
W(\phi=\pi/2)$ and $\hat W_z$ are generators of the SU(2) Lie
algebra. The uncertainty relation $\var{W_{x}}\var{W_{y}}\ge
(1/4)|\<\hat W_z\>|^2$, justifies defining difference squeezing by
Eq.~(\ref{N23}). One can find that
\begin{equation}
  \var{W_{\phi}} = \varn{W_{\phi}} + \frac14 (\<\hat n_1\>+\<\hat n_2\>).
\label{N25}
\end{equation}
By recalling Criterion~3 for $\F=(1,\hat W_{\phi})$, it is seen
that
\begin{eqnarray}
  d_{\F}^{\rm (n)} =\varn{W_{\phi}} \ncl 0, \label{N26}
\end{eqnarray}
in analogy to Eq.~(\ref{N22}). And the condition for sum
squeezing, given by Eq.~(\ref{N23}), can be formulated as:
\begin{equation}
  \dfn < - \frac12 \min_{i=1,2} \mean{\hat n_i}.
\label{z1}
\end{equation}
So, states exhibiting difference squeezing are nonclassical. But
also states satisfying
\begin{eqnarray}
  \frac14|\<\hat n_1\>-\<\hat n_2\>| \le  \var{W_{\phi}}
  < \frac14(\<\hat n_1\>+\<\hat n_2\>) \label{N27}
\end{eqnarray}
are nonclassical although {\em not} exhibiting difference
squeezing. The first inequality in Eq.~(\ref{N27}) corresponds to
condition opposite to squeezing condition given by
Eq.~(\ref{N23}).

Criterion~3 can also be applied to the multimode generalization of
difference squeezing, which can be defined via the
operator~\cite{An00}:
\begin{equation}
\hat {\cal W}_\phi = \frac12 {\rm e}^{-i \phi}\prod_{k=1}^K \hat
a_k \prod_{m=K+1}^M \hat a_m^\dagger + {\rm H.c.}  \label{z12}
\end{equation}
for any $K<M$. For simplicity, hereafter, we skip the limits of
multiplication in $\prod_{k}$ and $\prod_{m}$. The commutation
relation
\begin{equation}
[\hat {\cal W}_\phi ,\hat {\cal W}_{\phi+\pi/2} ]=\frac{i}2 \hat
C, \label{z13}
\end{equation}
where
\begin{equation}
\hat C=\prod_{k}(1+\hat n_k)\prod_{m}\hat n_m -\prod_{k}\hat n_k
\prod_{m} (1+\hat n_m), \label{z14}
\end{equation}
justifies the choice of the following condition for multimode
difference squeezing along the direction $\phi$~\cite{An00}:
\begin{equation}
\var{{\cal W}_\phi} < \frac{|\mean{\hat C}|}4. \label{z18}
\end{equation}
We find that
\begin{equation}
\var{{\cal W}_\phi}=\varn{{\cal W}_\phi}+ \frac{|\mean{\hat D}|}4,
\label{z15}
\end{equation}
where
\begin{equation}
\hat D=\prod_{k}(1+\hat n_k)\prod_{m}\hat n_m +\prod_{k}\hat n_k
\prod_{m} (1+\hat n_m)-2 \prod_{j=1}^M\hat n_j. \label{z17}
\end{equation}
By applying Criterion~3 for $\hat F=(1,\hat {\cal W}_\phi)$, we
obtain the following condition for multimode difference squeezing:
\begin{equation}
\dfn = \varn{{\cal W}_\phi} < \frac14 \left(|\mean{\hat
C}|-\mean{\hat D}\right), \label{z19}
\end{equation}
which corresponds to the original condition, given by
Eq.~(\ref{z18}). For states exhibiting difference squeezing, the
right-hand side of Eq.~(\ref{z19}) is negative. In fact, if
$\mean{\hat C}>0$ then
\begin{equation}
\hat C - \hat D = -2 \prod_{k}\hat n_k \left(\prod_{m}(1+\hat n_m)
-\prod_{m}\hat n_m  \right) < 0, \label{z20}
\end{equation}
otherwise
\begin{equation}
\hat C - \hat D = -2 \left(\prod_{k}(1+\hat n_k) -\prod_{k}\hat
n_k  \right) \prod_{m}\hat n_m   < 0. \label{z21}
\end{equation}
It is seen that the difference squeezing condition is stronger
than the nonclassicality condition $\dfn\ncl 0$. This means that
states satisfying inequalities
\begin{eqnarray}
  \frac14 \left(|\mean{\hat C}|-\mean{\hat D}\right) \le \varn{{\cal W}_\phi}  <  0
\label{z21a}
\end{eqnarray}
are nonclassical but {\em not} exhibiting difference squeezing.

%------------------------------------------------------------------
\section{Unified derivations of criteria for one-time photon-number correlations}

Various criteria for the existence of nonclassical photon-number
intermode phenomena in two-mode radiation fields have been
proposed (see, e.g.,
Refs.~\cite{Reid86,Agarwal88,Lee90,DodonovBook,VogelBook,MandelBook,PerinaBook}).
Here, we give a few examples of such nonclassical phenomena
revealed by single-time moments.

%------------------------------------------------------------------
\subsection{Sub-Poisson photon-number correlations}

The {\em squeezing} of the sum ($\hat n_+=\hat n_1 +\hat n_2$) or
difference ($\hat n_-=\hat n_1 -\hat n_2$) of photon numbers
occurs if
\begin{eqnarray}
 \varn{n_{\pm}}
 &<& 0, \label{N28}
\end{eqnarray}
which can be interpreted as the photon-number sum/difference {\em
sub-Poisson statistics}, respectively~\cite{PerinaBook}. These are
nonclassical effects, as can be seen by analyzing the
$P$~function:
\begin{equation}
  \varn{n_{\pm}} = \intda P(\bm{\alpha,\alpha}^*)
  [(|\alpha_1|^2\pm|\alpha_2|^2) -\<\hat n_{\pm}\>]^2, \label{N29}
\end{equation}
where $\bm{\alpha}=(\alpha_1,\alpha_2)$. Thus, photon-number
squeezing implies the nonpositivity of the $P$~function. The same
conclusion can also be drawn by applying Criterion~3 for
$\F_{\pm}=(1,\hat n_{\pm})$, which leads to
\begin{eqnarray}
  d_{\F_{\pm}}^{\rm (n)} = \DET{1&\mean{\hat n_{\pm}}}
  {\mean{\hat n_{\pm}}&\normal{\hat n_{\pm}^2}} = \varn{n_{\pm}} \ncl 0.
\label{N30}
\end{eqnarray}

%------------------------------------------------------------------
\subsection{Agarwal's nonclassicality criterion}

Here, we consider an example of the violation of the CSI for two
modes at the same evolution time. Other examples of violations of
the CSI for a single mode, but at two different evolution times,
are discussed in Appendix C in relation to photon antibunching and
hyperbunching.

By considering the violation of the following CSI:
\begin{eqnarray}
   \normal{\hat n_1^2}\normal{\hat n_2^2} &\cl& \< \hat n_1 \hat
   n_2\>^2,
\label{x15}
\end{eqnarray}
Agarwal~\cite{Agarwal88} introduced the following nonclassicality
parameter:
%-----------------------------------------------------------------
\begin{equation}
I_{12} = \frac{\sqrt{\langle :\hat{n}_1^2:\rangle  \langle
:\hat{n}_2^2:\rangle}} {\mean{\hat{n}_1 \hat{n}_2}}-1.
 \label{x14}
\end{equation}
Explicitly, the nonclassicality of phenomena described by a
negative value of $I_{12}$ is also implied by Criterion~3 for
$\F=(\hat n_1,\hat n_2)$, which results in
\begin{eqnarray}
  \dfn &=& \DET{\normal{\hat n_1^2} & \mean{\hat{n}_1 \hat{n}_2}}
  {\mean{\hat{n}_1 \hat{n}_2} & \normal{\hat n_2^2}} \ncl 0.
\label{x17}
\end{eqnarray}

%------------------------------------------------------------------
\subsection{Lee's nonclassicality criterion}

The Muirhead classical inequality~\cite{Muirhead} is a
generalization of the arithmetic-geometric mean inequality. Lee
has formulated this inequality as follows~\cite{Lee90}
\begin{equation}
D_{12} = \langle :\hat{n}_1^2:\rangle + \langle
:\hat{n}_2^2:\rangle - 2 \langle \hat{n}_1 \hat{n}_2\rangle  \cl
0. \label{x30}
\end{equation}
The nonclassicality of correlations with a negative value of the
parameter $D_{12}$ is readily seen by applying Criterion~3 for
$\F=(\hat n_1-\hat n_2)\equiv (\hat n_{-})$, which yields
\begin{equation}
   D_{12} = \normal{\hat{n}_{-}^2} \ncl 0.
   \label{x30a}
\end{equation}
For comparison, let us analyze Criterion~3 for $\F=(1,\hat
n_{-})$, which leads to
\begin{equation}
  \dfn = \normal{\hat{n}_{-}^2} -\mean{\hat{n}_{-}}^2 \cl 0.
  \label{x31}
\end{equation}
Clearly
\begin{equation}
  D_{12} < 0 \Rightarrow \dfn \ncl 0. \label{x31a}
\end{equation}
Thus, the criterion given by Eq.~(\ref{x31}) detects more
nonclassical states than that based on the $D_{12}$ parameter.

Alternatively, a direct application of the relation
\begin{equation}
  D_{12} = \intda P(\bm{\alpha,\alpha}^*)
  (|\alpha_1|^2- |\alpha_2|^2)^2 \ncl 0 \label{x91}
\end{equation}
also implies the nonpositivity of the $P$~function in some regions
of phase space.

%------------------------------------------------------------------
\section{Unified derivations of criteria for two-time photon-number
correlations}

Here, we consider the two-time single-mode photon-number
nonclassical correlations on examples of photon antibunching and
photon hyperbunching.

%------------------------------------------------------------------
\subsection{Photon antibunching}

The {\em photon
antibunching}~\cite{Kimble77,Walls79,Loudon80,VogelBook,MandelBook}
of a stationary or nonstationary single-mode field can be defined
via the two-time second-order intensity correlation function given
by
\begin{eqnarray}
G^{(2)}(t,t+\tau) &=& \normalo{\hat{n}(t)\hat{n}(t+\tau)} \notag\\
&=&\langle \hat{a}^{\dagger}(t)\hat{a}^{\dagger}(t+\tau)
\hat{a}(t+\tau)\hat{a}(t)\rangle\quad\quad \label{y01}
\end{eqnarray}
or its normalized intensity correlation functions defined as
\begin{equation}
g^{(2)}(t,t+\tau )= \frac{G^{(2)}(t,t+\tau )}{\sqrt{
G^{(2)}(t,t)G^{(2)}(t+\tau ,t+\tau )}}, \label{y02}
\end{equation}
where $\dd\,\dd$ denotes the time order and normal order of field
operators. Photon antibunching occurs if $g^{(2)}(t,t)$ is a
strict local minimum at $\tau =0$ for $g^{(2)}(t,t+\tau )$
considered as a function of $\tau$ (see, e.g.,
Refs.~\cite{MandelBook,Miranowicz99a}):
%----------------------------------------------------------------------
\begin{eqnarray}
g^{(2)}(t,t+\tau ) > g^{(2)}(t,t). \label{y05}
\end{eqnarray}
{\em Photon bunching} occurs if $g^{(2)}(t,t+\tau)$ decreases,
while {\em photon unbunching} appears if $g^{(2)}(t,t+\tau)$ is
locally constant.

For {\em stationary} fields [i.e., those satisfying
$G^{(2)}(t,t+\tau)=G^{(2)}(\tau)$ so
$g^{(2)}(t,t+\tau)=g^{(2)}(\tau)$], Eq.~(\ref{y05}) reduces to the
standard definition of photon
antibunching~\cite{VogelBook,MandelBook}:
\begin{eqnarray}
g^{(2)}(\tau )> g^{(2)}(0). \label{y05b}
\end{eqnarray}

Photon antibunching, defined by Eq.~(\ref{y05}), is a nonclassical
effect as it corresponds to the violation of the Cauchy-Schwarz
inequality:
%----------------------------------------------------------------------
\begin{eqnarray}
G^{(2)}(t,t)G^{(2)}(t+\tau ,t+\tau ) \cl \big[G^{(2)}(t,t+\tau
)\big]^2. \label{y09}
\end{eqnarray}
As shown in Ref.~\cite{Vogel08}, this property follows from
Criterion~3 based on the generalized definition of space-time
$P$~function, given by~(\ref{VogelP}). In fact, by assuming
$\F=(\hat n(t),\hat n(t+\tau))$, which leads to
\begin{eqnarray}
  \dfn &=& \DET{\normalo{\hat n^2(t)} &
  \normalo{\hat n(t)\hat n(t+\tau)}}
  {\normalo{\hat n(t)\hat n(t+\tau)} & \normalo{\hat n^2(t+\tau)}}
\notag \\
 &=& \DET{G^{(2)}(t,t) & G^{(2)}(t,t+\tau)}
  {G^{(2)}(t,t+\tau) & G^{(2)}(t+\tau,t+\tau)} \ncl 0.\quad\quad
\label{x23}
\end{eqnarray}

%------------------------------------------------------------------
\subsection{Photon hyperbunching}

{\em Photon hyperbunching}~\cite{Jakob01}, also referred to as
photon antibunching effect~\cite{Miranowicz99b}, can be defined
as:
%----------------------------------------------------------------------
\begin{eqnarray}
\overline{g}^{(2)}(t,t+\tau )> \overline{g}^{(2)}(t,t),
\label{y05a}
\end{eqnarray}
given in terms of the correlation coefficient~\cite{Berger93}
%----------------------------------------------------------------------
\begin{equation}
\overline{g}^{(2)}(t,t+\tau )=  \frac{\overline{G}^{(2)}(t,t+\tau
)}{\sqrt{ \overline{G}^{(2)}(t,t)\overline{G}^{(2)}(t+\tau ,t+\tau
)}}, \label{y07}
\end{equation}
where the covariance $\overline{G}^{(2)}(t,t+\tau)$ is given by
\begin{equation}
\overline{G}^{(2)}(t,t+\tau) = G^{(2)}(t,t+\tau) -G^{(1)}(t)
G^{(1)}(t+\tau), \label{y08}
\end{equation}
and $G^{(1)}(t)=\langle \hat n(t)\rangle =\langle \hat{a}^{\dagger
}(t) \hat{a}(t)\rangle$ is the light intensity. It is worth noting
that, for {\em stationary} fields, the definitions given by
Eqs.~(\ref{y05}) and~(\ref{y05a}) are equivalent and equivalent to
definitions of photon antibunching based on other normalized
correlation functions, e.g.,
\begin{equation}
\tilde g^{(2)}(t,t+\tau )=\frac{G^{(2)}(t,t+\tau )}{[ G^{(1)}(t)]
^{2}}. \label{y08a}
\end{equation}
However for {\em nonstationary} fields, these definitions
correspond in general to different photon antibunching
effects~\cite{Miranowicz99a,Miranowicz99b,Jakob01}.

Analogously to Eq.~(\ref{y05}), the photon hyperbunching, defined
by Eq.~(\ref{y05a}), can occur for nonclassical fields violating
the Cauchy-Schwarz inequality:
\begin{equation}
\overline{G}^{(2)}(t,t)\overline{G}^{(2)}(t+\tau ,t+\tau ) \cl
\big[\overline{G}^{(2)}(t,t+\tau )\big]^2. \label{y10}
\end{equation}
Again, the nonclassicality of this effect can be shown by applying
Criterion~3 for the space-time $P$~function, given
by~(\ref{VogelP}), assuming $\F=(\Delta \hat n(t),\Delta \hat
n(t+\tau))$, where $\Delta \hat n(t) =\hat n(t)-\mean{\hat n(t)}$.
Thus, one obtains
\begin{equation}
  \dfn = \DET{\overline G^{(2)}(t,t) & \overline G^{(2)}(t,t+\tau)}
  {\overline G^{(2)}(t,t+\tau) & \overline G^{(2)}(t+\tau,t+\tau)} \ncl
  0,
\label{x27}
\end{equation}
which is equivalent to Eq.~(\ref{y05a}). Alternatively, by
choosing $\hat F=(1,\hat n(t),\hat n(t+\tau))$, one finds
\begin{equation}
\dfn = \DETT {1&\mean{\hat n(t)}&\mean{\hat n(t+\tau)}}
{\mean{\hat n(t)}&\mean{\dd\hat n^2(t)\dd} &\mean{\dd \hat n(t)
\hat n(t+\tau)\dd}} {\mean{\hat n(t+\tau)}&\mean{\dd \hat n(t)
\hat n(t+\tau)\dd}& \mean{\dd\hat n^2(t+\tau)\dd}}, \label{z34}
\end{equation}
which is equal to the determinant given by Eq.~(\ref{x27}). By
comparing Eqs.~(\ref{x27}) and~(\ref{z34}), analogously to
Eqs.~(\ref{N18}) and~(\ref{z36}), it is seen the advantage of
using polynomial, instead of monomial, functions of moments in
$\F$.

Finally, it is worth noting that the {\em single-mode sub-Poisson}
photon-number statistics, defined by the condition $\varn{n}<0$,
although also referred to as {\em photon antibunching}, is an
effect different from those defined by Eqs.~(\ref{y05})
and~(\ref{y05a}), as shown by examples in Ref.~\cite{Zou90}.

\end{appendix}

%------------------------------------------------------------------

\end{document}